\documentclass[]{pasj01}
\Received{2019/05/31}%{yyyy/mm/dd}
\Accepted{2019/08/10}%{yyyy/mm/dd}
\usepackage{url}
\usepackage{color}
\usepackage{lscape}
\usepackage{longtable}
\begin{document} 

\title{ 
The whole picture of the large-scale structure of the CL1604 supercluster at z$\sim$0.9
}  

%%% begin:list of authors
% Do NOT capitalize all letters in "textsc".
\author{Masao \textsc{Hayashi}\altaffilmark{1}%
%\thanks{Example: Present Address is xxxxxxxxxx}
}
\email{masao.hayashi@nao.ac.jp}
\author{Yusei \textsc{Koyama}\altaffilmark{2,3}}
\author{Tadayuki \textsc{Kodama}\altaffilmark{4}}
\author{Yutaka \textsc{Komiyama}\altaffilmark{1}}
\author{Yen-Ting \textsc{Lin}\altaffilmark{5}}
\author{Satoshi \textsc{Miyazaki}\altaffilmark{1}}
\author{Rhythm \textsc{Shimakawa}\altaffilmark{2}}
\author{Tomoko L. \textsc{Suzuki}\altaffilmark{1,4}}
\author{Ichi \textsc{Tanaka}\altaffilmark{2}}
\author{Moegi \textsc{Yamamoto}\altaffilmark{3}}
\author{Naoaki \textsc{Yamamoto}\altaffilmark{4}}
\altaffiltext{1}{National Astronomical Observatory of Japan, 2-21-1 Osawa, Mitaka, Tokyo 181-8588, Japan}
\altaffiltext{2}{Subaru Telescope, National Astronomical Observatory of Japan, 650 North A'ohoku Place, Hilo, HI 96720, USA}
\altaffiltext{3}{Department of Astronomical Science, SOKENDAI (The Graduate University for Advanced Studies), Mitaka, Tokyo 181-8588, Japan}
\altaffiltext{4}{Astronomical Institute, Tohoku University, Aramaki, Aoba-ku, Sendai 980-8578, Japan}
\altaffiltext{5}{Academia Sinica Institute of Astronomy and Astrophysics, P.O. Box 23-141, Taipei 10617, Taiwan}
%%% end:list of authors

%% `\KeyWords{}' always has to be placed before ``\maketitle'' 
%%  List of Key Words:  https://academic.oup.com/pasj/pages/Pasj_Keywords 

\maketitle

\begin{abstract}
We present the large-scale structure over more than 50 comoving Mpc
scale at z $\sim$ 0.9 where the CL1604 supercluster, which is one of
the largest structures ever known at high redshifts, is embedded. 
The wide-field deep imaging survey by the Subaru Strategic Program with 
Hyper Suprime-Cam reveals that the already-known CL1604 supercluster
is a mere part of larger-scale structure extending to both the north
and the south. We confirm that there are galaxy clusters at three
slightly different redshifts in the northern and southern sides of the
supercluster by determining the redshifts of 55 red-sequence galaxies
and 82 star-forming galaxies in total by the follow-up spectroscopy with
Subaru/FOCAS and Gemini-N/GMOS. This suggests that the structure ever
known as the CL1604 supercluster is the tip of the iceberg. We
investigate stellar population of the red-sequence galaxies using 4000
\AA\ break and Balmer H$\delta$ absorption line. Almost all of the
red-sequence galaxies brighter than 21.5 mag in $z$-band show an old
stellar population with $\gtrsim2$ Gyr. The comparison of composite
spectra of the red-sequence galaxies in the individual clusters show
that the galaxies at a similar redshift have similar stellar
population age, even if they are located $\sim$50 Mpc apart from each
other. However, there could be a large variation in the star formation
history. Therefore, it is likely that galaxies associated with the
large-scale structure at 50 Mpc scale formed at almost the same time,
have assembled into the denser regions, and then have evolved with
different star formation history along the hierarchical growth of the
cosmic web.     
\end{abstract}

\section{Introduction}
The canonical model of the cold dark matter structure formation
suggests that small-scale building blocks merge via gravitational
interactions and then grow into more massive structures in the universe. 
Distribution of dark matter deviates from uniformity as
time goes on, and large-scale structures of dark matter haloes that
consist of voids, filaments, groups, and clusters become prominent
gradually (e.g., \cite{Springel2005,Vogelsberger2014,Schaye2015}).   
Given that galaxies are hosted in dark matter haloes, galaxy formation
and evolution is expected to be closely linked to the growth of
large-scale structures.  
Therefore, it is important to understand how galaxies have evolved
while the dark matter haloes grow hierarchically. The cosmological
simulations predict that the local clusters consist of galaxies that
were located in the wide areas over several tens Mpc scale at high
redshifts and then assembled into the high-density regions, although
the cluster member galaxies are currently bounded within about Mpc
scale \citep{Muldrew2015,Chiang2017}. This suggests that when we
investigate the large-scale structures at higher redshifts, a wider
survey volume is essential to reveal the structure formation.    

Observationally, the cosmic web structures of galaxies at z $\lesssim$
1 are revealed by wide-field spectroscopic surveys such as  
Sloan Digital Sky Survey (SDSS, \cite{Blanton2003,Abazajian2009}), 
2dF Galaxy Redshift Survey (2dFGRS, \cite{Colless2001}), 
VIMOS VLT Deep Survey (VVDS, \cite{Garilli2008}),
VIMOS Public Extragalactic Redshift Survey (VIPERS, \cite{Guzzo2014}),
Observations of Redshift Evolution in Large-Scale Environments (ORELSE) Survey \citep{Lubin2009,Gal2008},
DEEP2 survey \citep{Gerke2012,Newman2013},
Prism multi-object survey (PRIMUS, \cite{Skibba2014,Coil2011}),
and HectoMAP survey \citep{Hwang2016,Sohn2018}. 
Indeed, as seen in cosmological simulation of the cold dark matter
model, it is observed that galaxy clusters are located at the intersections of
filamentary structures over the several tens Mpc scale and galaxy
clusters also compose larger-scale superclusters embedded in the
cosmic web \citep{Haynes1986,Nakata2005,Swinbank2007,Tanaka2009a,Einasto.M2011,Liivamagi2012,Chow-Martinez2014,Tully2014,Nevalainen2015,Kim2016,Lietzen2016,Pompei2016,Kraan-Korteweg2017,Haines2018,Paulino-Afonso2018,Koyama2018}.
As represented by the large-scale structures such as 
the XLSSsC N01 supercluster at $z\sim0.3$ \citep{Guglielmo2018}, 
the BOSS Great Wall at $z\sim0.47$ \citep{Lietzen2016}, 
the Cl J021734-0513 supercluster at $z\sim0.65$ \citep{Galametz2018},
the COSMOS Wall at $z\sim0.73$ \citep{Iovino2016}, 
the RCS 2319+00 supercluster \citep{Gilbank2008},
and the Lynx supercluster at $z\sim1.3$ \citep{Nakata2005},
there are many known structures at high redshifts and they are
interesting and important from the perspective of the growth of the
structures in the universe.  
Among the high-$z$ structures surveyed, the CL1604 supercluster at
$z\sim0.9$ investigated by the ORELSE survey is one of the largest
complex structures ever known at high redshifts over $\sim 0.5 \deg$
($\sim26$ comoving Mpc), which consists of three galaxy clusters with 
$M_{cl}>10^{14} M_{\solar}$ and at least five galaxy groups with $M_{g}>10^{13} M_{\solar}$
\citep{Lemaux2012,Hung2019}. 

\begin{figure*}
 \begin{center}
   \includegraphics[width=\linewidth,trim=65 0 20 20,clip]{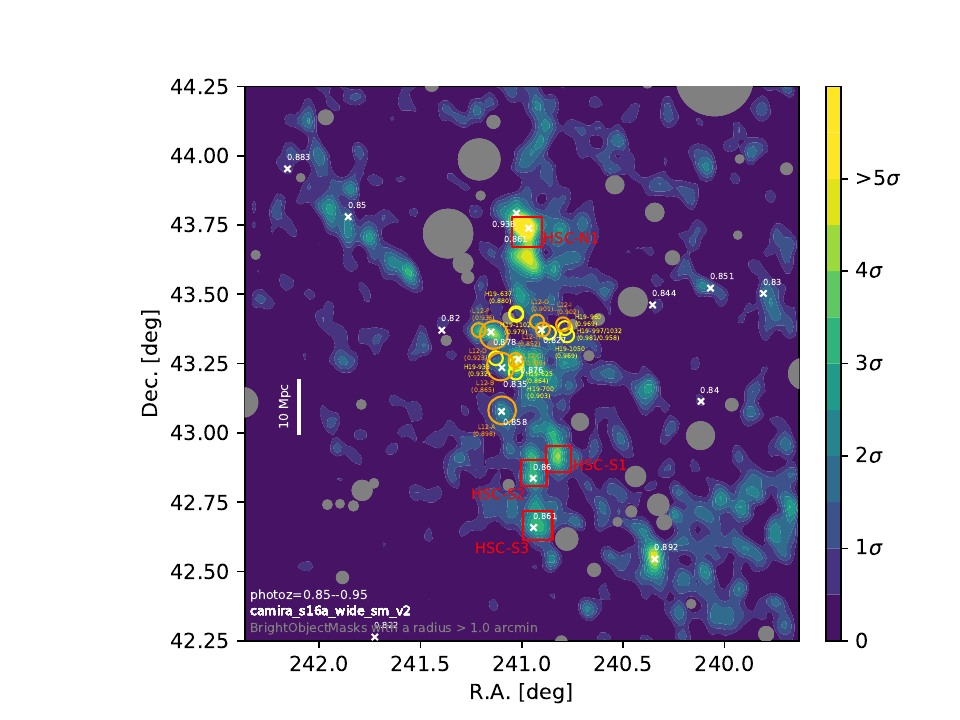}
 \end{center}
 \caption{Distribution of galaxies at $z\sim0.9$ is shown using a
   kernel-density estimate with a Gaussian kernel. The galaxies are
   selected from the HSC-SSP S16A data by applying the criteria of
   photometric redshifts ({\tt ephor\_ab}, \cite{HSCSSPDR1PHOTOZ}) of
   0.85--0.95. The contours are drawn on the grid of 1$\times$1
   arcmin$^2$ and a bandwidth two times larger than the bin size is
   used in the Gaussian kernel. The sigma is estimated from the
   standard deviation of the densities in the bins. The bright object
   masks with a radius larger than 1 arcmin are shown in gray
   \citep{Coupon18HSC}. The galaxy clusters and groups 
   with $>10^{13.5} M_{\solar}$ at redshifts of 0.8--1.0 reported by 
   \citet{Lemaux2012} and \citet{Hung2019} are shown
   by the orange and yellow 
   circles with the redshift
   information. The white crosses are galaxy clusters selected on the
   HSC-SSP S16A data by the red-sequence finder named CAMIRA
   \citep{Oguri2014,Oguri2018}. The white numbers are the redshifts of
   the clusters estimated by CAMIRA. The overdense regions of galaxies
   selected based on the photometric redshifts traces both the
   clusters/groups ever known and the CAMIRA clusters.
   The red squares are overdense regions where the spectroscopic
   follow-up observations are conducted by this study.
 }\label{fig:map}
\end{figure*}

At present, there are wide-field imaging surveys covering over several
hundreds to about a thousand deg$^2$ such as Pan-STARRS \citep{Chambers2016}, 
DES \citep{DES}, and KiDS \citep{deJong2015}. 
Among them, Subaru Strategic Program (SSP) with Hyper
Suprime-Cam (HSC, \cite{Miyazaki18HSC,Komiyama18HSC,Kawanomoto18HSC,Furusawa18HSC})  
is an ongoing imaging survey in $grizy$ five bands that combines both
a coverage of 1400 deg$^2$ and a depth of $r_{AB}\approx26$ by making
full use of a field of view (FoV) of HSC, 1.77 deg$^2$, and 8-m class
Subaru telescope \citep{HSCSSP}. The HSC-SSP survey provides us with
wide-field, deep data set that is essential to investigating large-scale
structures at high redshifts. The survey area of the HSC-SSP covers the
regions including the already-known structures of the CL1604
supercluster. The HSC data in five broadband of $grizy$ became
available in the DR1 S16A internal release in August 2016
\citep{HSCSSPDR1}. 
Then, the second public data release (PDR2) of HSC-SSP 
data\footnote{\url{https://hsc.mtk.nao.ac.jp/ssp/}\label{foot:hscsspwebsite}}  
is in May 2019 \citep{HSCSSPDR2}. 
Interestingly, we have found a strong evidence that the already-known
CL1604 supercluster is embedded within more extended large-scale
structures (Fig.~\ref{fig:map}). 
We are on the verge of revealing a {\it complete picture} of the
CL1604 supercluster with more than 50 comoving Mpc scale ($>1$deg) at $z\sim0.9$.

This paper aims to reveal the whole picture of the large-scale
structures around the CL1604 by the HSC-SSP imaging data and confirm
them spectroscopically, which is described in section
\ref{sec:obsanddata}. This undoubtedly allows us to investigate how
the galaxies associated with the large-scale structures have assembled
into the denser regions and then evolved along the hierarchical growth
of the cosmic web of the underlying dark matter haloes. 
In section \ref{sec:discussions}, we investigate the stellar
populations of red-sequence galaxies,  
compare the stellar populations between the galaxies in different
clusters confirmed and discuss the formation process of the
large-scale structure.  
Finally, our conclusions are presented in section
\ref{sec:conclusions}. Throughout this paper, 
magnitudes are presented in the AB system \citep{Oke1983}. The
cosmological parameters of 
H$_0$ = 70 km s$^{-1}$ Mpc$^{-1}$, $\Omega_m$ = 0.3, and $\Omega_\Lambda$ = 0.7  
are adopted.

\section{Revealing the whole picture of the CL1604 supercluster}
\label{sec:obsanddata}

\subsection{Wide-field imaging survey with Subaru HSC}
\label{sec:LssByHscimag}

The HSC data around the CL1604 supercluster are available from both
the HSC-SSP S16A internal release and PDR2. 
Among three layers of the
HSC-SSP survey of Wide, Deep, and UltraDeep, the CL1604 supercluster
is located in the Wide layer with a field name of HECTOMAP
\citep{HSCSSPDR2}. The HECTOMAP field is designed to cover the field
of \timeform{13h20m} $\leqq$ R.A.~$\leqq$ \timeform{16h40m} and
\timeform{42D} $\leqq$ Dec.~$\leqq$ \timeform{44.5D}, i.e. $\sim$
90 deg$^2$ \citep{HSCSSP}.
In this paper, the limited $\sim2\times2$ deg$^2$ area is used to
reveal the large-scale structures around the CL1604 supercluster.
The HSC-SSP is an on-going survey and the pipeline used for data
processing is also still developing \citep{HSCSSPDR2}.
The different version of data release means that the data are
processed by the different version of the pipeline. The S16A data are
processed with the pipeline, {\tt hscPipe v4.0.2} \citep{Bosch18HSC},
and the PDR2 data are processed with the latest pipeline, {\tt hscPipe v6} 
\citep{Bosch18HSC,HSCSSPDR2}.  
Measurement flags given by a version of the pipeline to each object,
which are used for e.g., object selection, validation of data quality
and reliability of photometry, cannot be necessarily coincident with
those by another version of the pipeline. Therefore, HSC data used in
this section are extracted from the S16A internal release
\citep{HSCSSPDR1}, not PDR2, to keep a consistency between our 
original investigation of the galaxy distribution around the CL1604
supercluster and the subsequent target selection for the follow-up
spectroscopy.  
A composite model magnitude named {\tt cmodel} is used for the
photometry of galaxies. The {\tt cmodel} photometry measures fluxes of
objects by simultaneously fitting two components of a de Vaucouleur
and an exponential profile convolved with a point-spread function
(PSF) (see \cite{Bosch18HSC} for more details). 

\begin{figure*}
 \begin{center}
   \includegraphics[width=\linewidth]{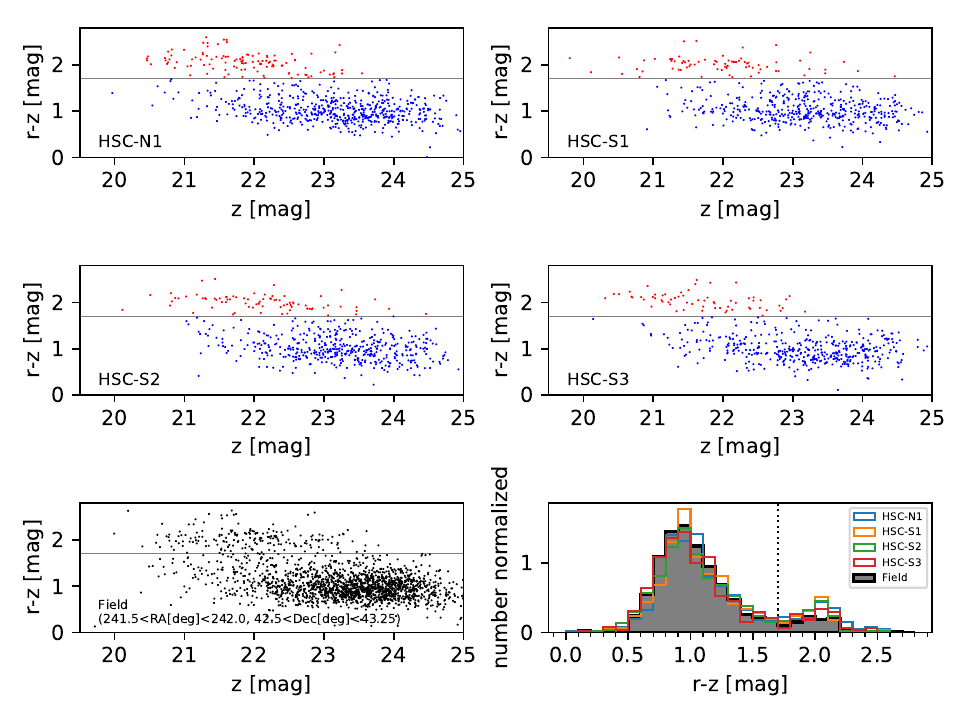}
 \end{center}
 \caption{Color -- magnitude diagram showing $r-z$ colors as a
   function of $z$ magnitude in the four regions where our follow-up
   spectroscopy was conducted. The galaxies with $r-z>1.7$ are classified as
   red-sequence galaxies, while the galaxies with $r-z<1.7$ are
   classified as star-forming galaxies. For the comparison, we
   arbitrarily define the field as a region with 241.5$<$R.A.$<242.0$
   and 42.5$<$Dec.$<$43.5. In the lower right panel, the normalized
   histograms of $r-z$ colors in each region are shown. The dashed
   line show the color of $r-z=1.7$. Compared with the field, the
   overdense regions show the larger number of the red-sequence
   galaxies, which can be visible by prominent red sequence.}\label{fig:CMD} 
\end{figure*}

The number density map of the photo-$z$ selected galaxies with
$z_{ph}=$ 0.85--0.95 \citep{HSCSSPDR1,HSCSSPDR1PHOTOZ} shows the
large-scale structures extending in north-to-south direction well
beyond the already-known structures of the CL1604 supercluster
(figure~\ref{fig:map}). In the S16A release, the photometric redshifts
calculated by six different codes are available. Among them, we use
the photo-$z$ calculated by the code named {\tt ephor\_ab} and the
photo-$z$ quality is $\sigma_z\sim0.035$ \citep{HSCSSPDR1PHOTOZ}. 
Note that even if we use the photo-$z$ calculated by any of the other
codes, the number density map of galaxies with $z_{ph}=$ 0.85--0.95 is
consistent with each other.  
As shown in figure~\ref{fig:map}, the structure of the photo-$z$
selected galaxies coincides with the distribution of the galaxy
clusters at $z_{ph,cl}$=0.9$\pm$0.1 ($\sigma_{z_{ph,cl}}\sim0.01$) identified
by a red sequence galaxy finder named CAMIRA \citep{Oguri2014,Oguri2018},
suggesting that we are seeing the {\it real} global structures at
$z\sim0.9$. 
We also make sure that galaxies with photometric redshifts of
0.70--0.80 or 0.95--1.05 do not show the similar structure of
overdensity regions and structures of galaxies with photometric
redshifts of 0.75--0.85 or 0.90--1.00 are not so prominent as shown
in figure~\ref{fig:map}. These support that the large-scale structure
of galaxies with photometric redshifts of 0.85--0.95 is real.  
Figure~\ref{fig:CMD} shows the color--magnitude diagram of the galaxies
selected in the several overdense regions, where the prominent red
sequence is seen compared with that in the general fields. Note that
we define the boundary of $r-z=1.7$ to distinguish red-sequence
galaxies from star-forming galaxies (figure~\ref{fig:CMD}). The
concentration of red galaxies also convinces us that the galaxies
selected are associated with the large-scale structures including the
CL1604 supercluster.  

Figure~\ref{fig:map} demonstrates that the Subaru HSC has the unique
ability to survey the large-scale structures at high redshifts thanks
to both the wide-field coverage and the depth of the data.
However, these evidences are all based on the photometric data. 
The spectroscopic confirmation is essential to proceed with further
detailed studies on the structure formation and cluster galaxy
evolution. To do that, we visually select four regions showing the
highest number density based on figure~\ref{fig:map}; one is to the
north of the already-known CL1604 supercluster and three are to the
south.     

\subsection{Spectroscopic confirmation of the large-scale structure}
\label{sec:MOSobs}

\begin{table*}
\tbl{Summary of the MOS observations.}{%
\begin{tabular}{cccccccccc}
\hline
mask & cluster & observed\footnotemark[$*$] & confirmed\footnotemark[$*$] & telescope & spectrograph & grating & integration & seeing & program ID \\
 & & red/blue\footnotemark[$\dag$] & red/blue\footnotemark[$\dag$] &  &  &  & [min] & [arcsec] & \\
\hline
1& HSC-N1& 15/14& 15/14& Subaru& FOCAS& VPH850& 105& 0.72--0.93& S18A-125 \\
2& HSC-N1& 16/22& 11/14& Gemini-N& GMOS& R400& 72.3& 0.74--0.90& GN-2018A-FT-107 \\
3& HSC-S1& 10/21& 10/16& Gemini-N& GMOS& R400& 72.3& 0.77--0.89& GN-2018A-FT-107 \\
4& HSC-S2& 12/22& 12/17& Gemini-N& GMOS& R400& 72.3& 0.60--0.68& GN-2018A-FT-107 \\
5& HSC-S3&  7/25&  7/21& Gemini-N& GMOS& R400& 72.3& 0.61--0.69& GN-2018A-FT-107 \\
\hline
\end{tabular}
}\label{tbl:obssummary}
\begin{tabnote}
\footnotemark[$*$] The number of galaxies that are observed or confirmed by the observations.\\
\footnotemark[$\dag$] The galaxies with $r-z>1.7$ are classified as red galaxies, and the others are blue galaxies (see also figure~\ref{fig:CMD}).
\end{tabnote}
\end{table*}

\subsubsection{Spectroscopy with Subaru and Gemini-N}
\label{sec:MOSobsSubaruGemini}

We used FOCAS/Subaru \citep{Kashikawa2002} and GMOS-North/Gemini
\citep{Hook2004} to conduct follow-up observations for galaxies
selected in section~\ref{sec:LssByHscimag} and then confirm
four overdense regions associated with the extended large-scale
structures at $z\sim0.9$ (figure~\ref{fig:map}). 
The observations were conducted through two programs of S18A-125 for
the FOCAS run (PI: T.~Kodama) and GN-2018A-FT-107 for the GMOS run
(PI: M.~Hayashi). The FOCAS run was in classical mode on March 8, 2018
(UT) and the GMOS run was in queue mode on June 14 and 16, 2018
(UT). The multi-object spectroscopy (MOS) observations are summarized
in table~\ref{tbl:obssummary}.   

FOCAS has a circular field of view (FoV) of 6 arcmin in diameter.
We used a MOS mask in the FOCAS run to observe the galaxies in the N1
region. 
Among the candidates within the FoV, red-sequence galaxies
with $z-$band mag $<$ 22.5 and $r-z=$ 1.8--2.3 are selected
as a first priority and then  
star-forming galaxies with $r-z<$ 1.6 are selected. 
Higher priority is given to brighter galaxies. In the mask,
28 slits are allocated to 29 galaxies, i.e., a pair of close targets
is observed with a single slit (15 red-sequence galaxies and 14
star-forming galaxies).  
A slit is allocated to a star for monitoring the sky condition such as
seeing and transparency. The slit width is 0.8 arcsec. We use VPH850
grating and SO58 order sorting filter, which provides spectral resolution of
$R=750$ for 0.8 arcsec slit and wavelength coverage of
5800--10350 \AA. Since 2 pixels are binned along spatial direction in
each readout of frame, a pixel scale is 1.19 \AA\ and 0.2076 arcsec per
pixel. The spectra in the sequence of the exposures were obtained by
dithering by $\pm1$ arcsec from the central position along the slit. We took
seven exposures with 900 sec on-source integration each, and thus the
total integration time is 1.75 hours. The seeing ranges between 0.72
and 0.93 arcsec, which are measured from the spectra of the monitoring
star. The sky was in almost photometric condition during the
observation with this mask.    

GMOS has a rectangular FoV of 5.5$\times$5.5 arcmin$^2$.
We used a MOS mask in the GMOS run to observe the galaxies in each
candidate of galaxy cluster, i.e., 4 MOS masks were used in total. 
We selected the target galaxies based on the same strategy as in the
FOCAS run. 
We targeted red-sequence galaxies with $z-$band mag $<$ 22.3 and
$r-z=$ 1.85--2.3 and star-forming galaxies with $r-z<$ 1.45. 
As shown in table~\ref{tbl:obssummary}, 31--38 slits are
allocated to red-sequence galaxies and star-forming
galaxies. Furthermore, a slit is allocated to a star in each mask for
monitoring the sky condition. The slit width is 1.0 arcsec. We use
R400 grating and OG515 blocking filter, which provides spectral
resolution of $R=959$ for 1.0 arcsec slit and wavelength coverage of
$>$6150 \AA. Since 2 pixels are binned along both spatial and spectral
direction in each readout of frame, a pixel scale is 1.52 \AA\ and
0.1614 arcsec per pixel. The spectra were obtained by dithering along
spectral direction at three central wavelengths of 7800, 7900, and
8000 \AA\ to fill the gap between the detectors. Note that we did not
use the Nod \& Shuffle mode. Five exposures with 868 sec on-source
integration each were taken under the dark night condition of cloud
cover CC=50\% (clear) and image quality IQ=70 (good), and thus the
total integration time is 1.21 hours. The seeing ranges between 0.60
and 0.90 arcsec, which are measured from the spectra of the monitoring
star.     

Consequently, we observed 164 galaxies (60 red-sequence galaxies and
104 star-forming galaxies) in the north FoV (HSC-N1) and the south
FoVs (HSC-S1, HSC-S2, and HSC-S3) in total
(see also table~\ref{tbl:obssummary}). Among the selected galaxies in 
each FoV, 58\%, 50\%, 44\%, and 32\% of the red-sequence galaxies were
observed in the N1, S1, S2, and S3 regions, respectively. For
star-forming galaxies, 16\%, 17\%, 21\%, and 16\% of the selected
galaxies were observed in the N1, S1, S2, and S3 regions,
respectively. Although we targeted almost the same fraction of
star-forming galaxies in each FoV, the fraction of red-sequence
galaxies observed in the S3 FoV is smaller than those in the other
FoVs. 

\subsubsection{Reduction of the spectroscopic data}

We used the {\tt FOCASRED} package, which includes the {\tt IRAF}% 
\footnote{IRAF is distributed by the National Optical Astronomy
  Observatory, which is operated by the Association of Universities
  for Research in Astronomy (AURA) under a cooperative agreement with
  the National Science Foundation.\label{foot:iraf}}
scripts, for the Subaru/FOCAS data reduction. 
We also used the Gemini {\tt IRAF}\footnotemark[\ref{foot:iraf}] package
and reduced the Gemini-N/GMOS data with {\tt PyRAF}%
\footnote{PyRAF is a product of the Space Telescope Science Institute,
  which is operated by AURA for NASA.} 
according to the GMOS Data Reduction Cookbook hosted on the US
National Gemini Office (NGO) pages\footnote{\url{http://ast.noao.edu/sites/default/files/GMOS_Cookbook/}}. 

We reduced both the FOCAS and GMOS data in a standard manner. 
All of bias subtraction, overscan subtraction, flat fielding,
distortion correction, wavelength calibration, sky-subtraction, coadd,
and flux calibration are conducted in order.
The wavelength calibration is based on the bright sky emission lines for
FOCAS data and the CuAr spectrum for GMOS data. As a spectroscopic
standard star, HZ44 and Feige66 are observed with FOCAS and GMOS,
respectively. The spectra of the standard stars are used to
correct for telluric absorption. Furthermore, to correct for the flux
loss from the slit, we normalize the flux density of the spectrum to
the broadband photometry for the red-sequence galaxies with stellar
continuum detected. We perform no correction to the star-forming
galaxies only detected in emission line, because we do not use the
fluxes of emission lines in this paper.   

\subsubsection{Redshift measurement}
\label{sec:redshift_measurement}

\begin{figure}
 \begin{center}
   \includegraphics[width=0.8\linewidth]{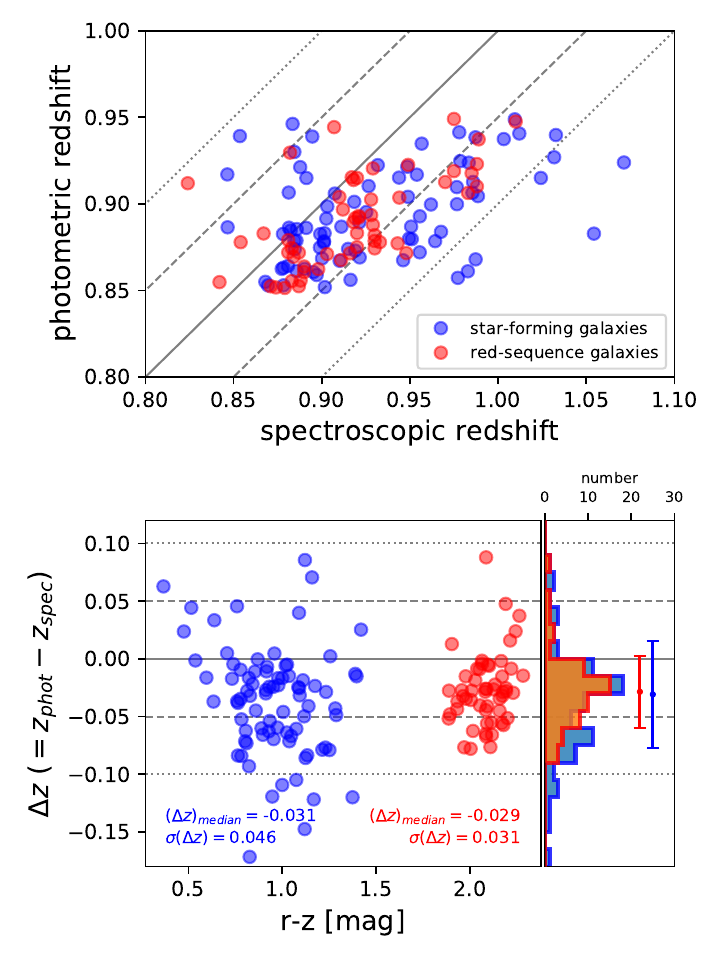}
 \end{center}
 \caption{The spectroscopic redshifts are compared with the
   {\tt ephor\_ab} photometric redshifts in the upper panel. The red-sequence galaxies
   are shown by red symbols, while star-forming galaxies are shown by
   blue symbols. The solid, dashed, and dotted line shows the
   difference of the two redshifts by $\pm$0.0, $\pm$0.05, and
   $\pm$0.1. In the lower panels, the difference between photometric
   and spectroscopic redshifts is shown as a function of the $r-z$ 
   color of galaxies. The histograms on the right-hand side show the
   distribution of the redshift difference in red-sequence (red) and
   star-forming (blue) galaxies. The dot and bar are the median and
   dispersion of the redshift difference.}\label{fig:zspecVSzph} 
\end{figure}

The redshifts for red-sequence galaxies are measured based on the
stellar continuum of the observed spectra as well as HSC broadband
photometry.   
For the redshift measurement, the optical broadband photometry of the
galaxies is extracted from the latest database of HSC-SSP PDR2
\citep{HSCSSPDR2}. 
The measurement values for the individual galaxies are shown in
Appendix~\ref{sec:appendix_Catalog}.   
If required, readers can obtain additional information of the galaxies
provided by the HSC-SSP survey in the data release
website\footnotemark[\ref{foot:hscsspwebsite}].   

We use the C++ version\footnote{\url{https://github.com/cschreib/fastpp}} 
of the FAST code \citep{Kriek2009} to fit the model spectral energy
distribution (SED) derived by stellar population synthesis to the
observed stellar continuum, where the redshift is one of the free parameters.
The model SED templates are generated by the code of \citet{bc03} and the
\citet{Chabrier2003} initial mass function (IMF) is assumed. The star
formation histories of the exponentially declining model are adopted,
where we set an e-folding time of $\mathrm{log}(\tau /\mathrm{yr})=8.5\mbox{--}10.0$ 
with ${\rm{\Delta }}\mathrm{log}(\tau /\mathrm{yr})\,=0.1$ \citep{Wuyts2011}.  
Ages of 0.1-–10.0 Gyr are acceptable with a step of 
${\rm{\Delta }}\mathrm{log}(\mathrm{age}/\mathrm{yr})=0.1$. The
extinction curve of \citet{Calzetti2000} is assumed, and $A_V$ ranges
from 0.0 to 3.0. Metallicity is fixed to the solar value. 
The absorption lines of Ca H+K and Balmer series are well fitted by
the model SED and useful to determine a redshift.
The uncertainty of the redshifts is $\Delta z$=0.001.
Among the 60 red-sequence galaxies observed, we obtain the redshifts for 55
galaxies with the stellar continuum detected with enough
signal-to-noise ratio. A success rate of the confirmation for
red-sequence galaxies is 91.7\%.  The redshifts for red-sequence galaxies
confirmed are listed in table~\ref{tbl:RSGs}.  

The redshifts for star-forming galaxies are determined based on
[O\emissiontype{II}]$\lambda3727$ emission line. Among the 104
star-forming galaxies observed, emission lines are detected from 82
galaxies. The detection of emission lines is visually inspected on
both 1D and 2D spectra for the individual galaxies. A success rate of
the confirmation for star-forming galaxies is 78.8\%. Although 28
galaxies have a single line detected in the spectrum, 54 galaxies have
multiple emission lines such as H$\beta$ and/or
[O\emissiontype{III}]$\lambda\lambda4959,5007$ detected. In the case
of the single line detected in the individual spectrum, we assume that
the emission line is [O\emissiontype{II}]$\lambda3727$. The assumption
should be reasonable, because the lines are detected around $\sim$7000
\AA\ in the observed frame and the wavelength of the line detection is
consistent with the expectation from the photometric redshifts
estimated with the optical broadband photometry. Note that the spectra
of 9 galaxies with the only single line detected do not cover the
wavelength of H$\beta$ and/or [O\emissiontype{III}]$\lambda\lambda4959,5007$.  
The redshifts of the galaxies with multiple emission lines detected
are more securely determined. The redshifts are measured from the peak
wavelength of the profile of [O\emissiontype{II}] emission line
derived by Gaussian fitting. The redshifts and the emission lines
detected for star-forming galaxies confirmed are listed in
table~\ref{tbl:SFGs}.     

\begin{figure}
 \begin{center}
   \includegraphics[width=\linewidth]{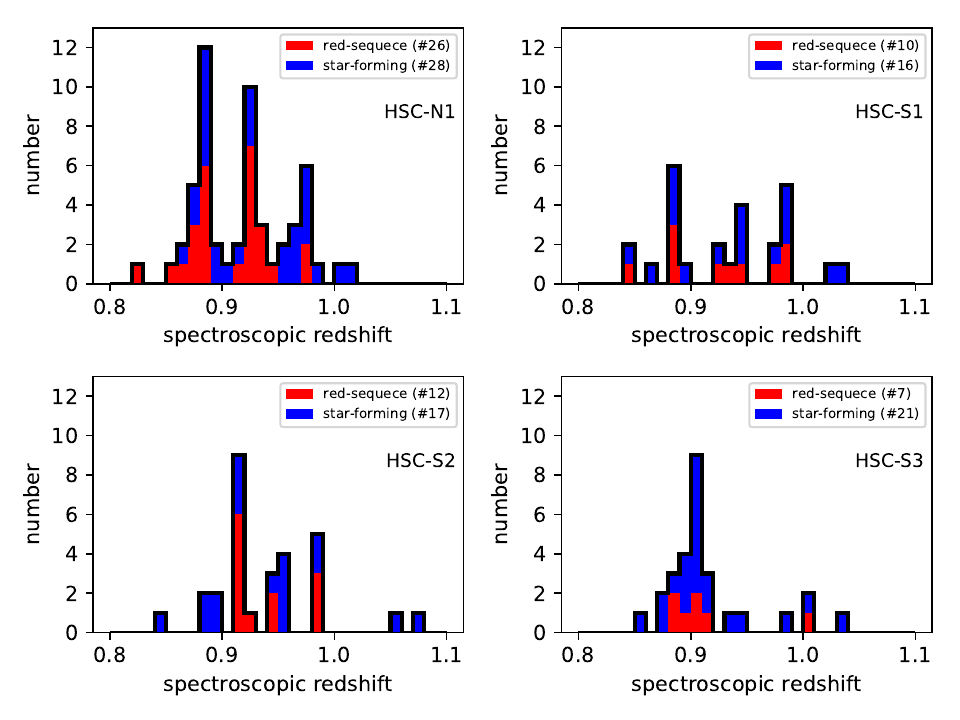}
 \end{center}
 \caption{Distribution of redshifts spectroscopically confirmed in
   each overdense region. The red histograms show the redshift
   distribution of red-sequence galaxies, the blue ones show those
   of star-forming galaxies. Note that the red histogram is stacked
   with the blue histogram, and the solid black histogram shows the
   total number of galaxies including red-sequence and star-forming
   galaxies. The number of the galaxies is shown in the
   legend.}\label{fig:dis_zspec}   
\end{figure}

\begin{figure*}
 \begin{center}
   \includegraphics[width=\linewidth,trim=15 0 0 0,clip]{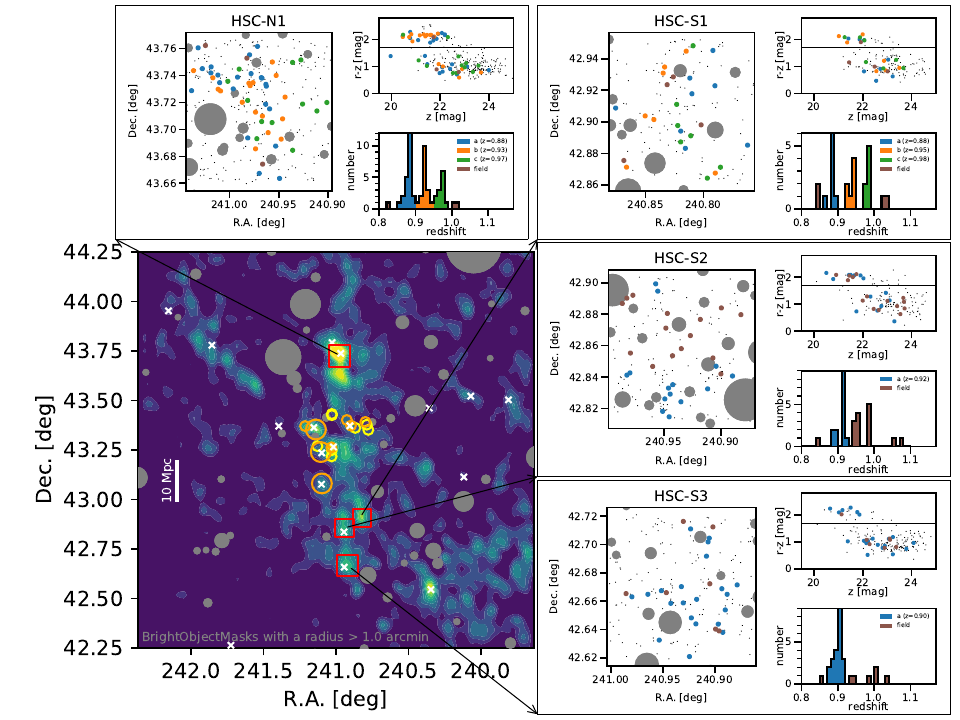}
 \end{center}
 \caption{This is an overview of the structures that we confirmed by
   this study. The map shown at the lower left is the same as
   figure~\ref{fig:map}. In each panels of upper left, upper right,
   middle right, and lower right, zoom-up view of the spatial
   distribution of the galaxies, the color--magnitude diagram, and the
   redshift distribution are shown in the four individual regions
   surveyed by this study. Note that each red square in the whole map
   corresponds to the field size of the zoom-up view and all of the
   bright object masks are shown in the zoom-up distribution. The
   galaxies spectroscopically confirmed are shown with the colors
   according to the redshifts. The black dots are galaxies selected
   with the photometric redshifts located in the zoom-up region.}\label{fig:map_summary}
\end{figure*}

\begin{figure}
 \begin{center}
   \includegraphics[width=\linewidth,trim=20 0 20 0,clip]{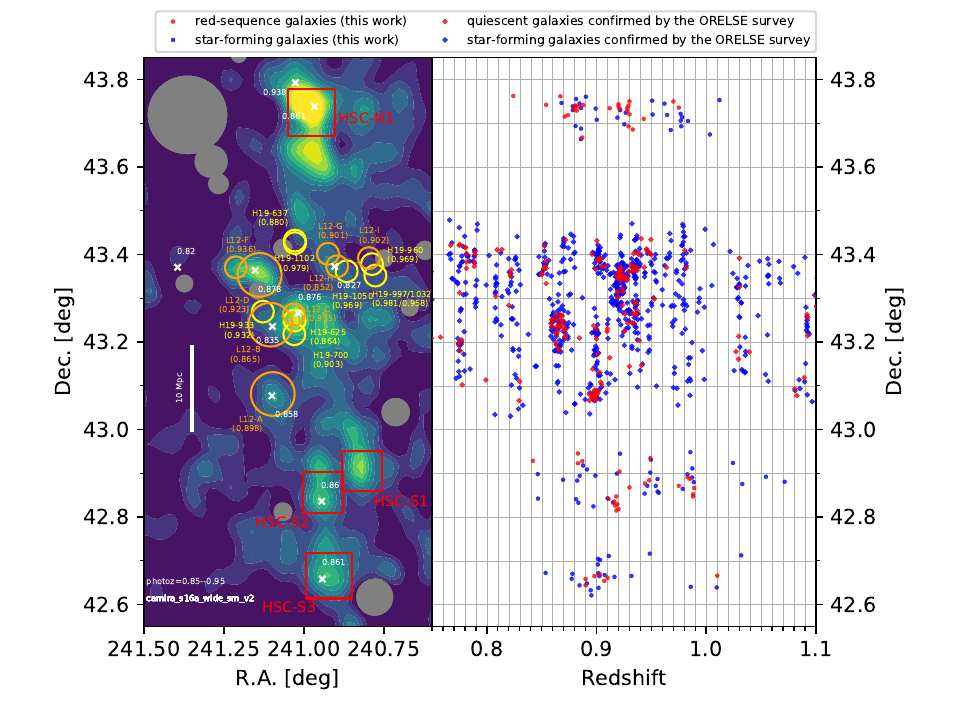}
 \end{center}
 \caption{
   The distribution of the confirmed galaxies in the
   redshift--declination space (right panel). The red (blue) circles
   are red-sequence (star-forming) galaxies confirmed by this work,
   while the red (blue) diamonds are quiescent (star-forming) galaxies
   confirmed by the ORELSE survey \citep{Lemaux2012,Lemaux2018}. 
   In the ORELSE survey, the quiescent galaxies are classified based
   on the rest-frame $NUV-r$ and $r-J$ colors whose criteria are
   defined by \citet{Lemaux2014}. In the left panel, as a reference,
   the same figure as figure~\ref{fig:map} is shown.  
 }\label{fig:specz_declination}
\end{figure}

\begin{table*}
\tbl{Summary of the structures confirmed.}{%
\begin{tabular}{lccccccccccc}
\hline
 & \multicolumn{2}{c}{HSC-N1}&& \multicolumn{2}{c}{HSC-S1}&& \multicolumn{2}{c}{HSC-S2}&& \multicolumn{2}{c}{HSC-S3} \\
 & \multicolumn{2}{c}{(\timeform{16h03m54.297s},}&& \multicolumn{2}{c}{(\timeform{16h03m16.652s},}&& \multicolumn{2}{c}{(\timeform{16h03m45.784s},}&& \multicolumn{2}{c}{(\timeform{16h03m41.025s},}\\  & \multicolumn{2}{c}{\timeform{+43D43'28.604''})}&& \multicolumn{2}{c}{\timeform{+42D54'18.768''})}&& \multicolumn{2}{c}{\timeform{+42D51'20.389''})}&& \multicolumn{2}{c}{\timeform{+42D39'58.111''})}\\ \cline{2-3}
\cline{5-6}
\cline{8-9}
\cline{11-12}
 & $\langle z \rangle$ & N && $\langle z \rangle$ & N && $\langle z \rangle$ & N && $\langle z \rangle$ & N\\
\hline
a ........ &
0.881&
22&&
0.884&
8&&
0.915&
14&&
0.901&
21\\
b ........ &
0.927&
17&&
0.946&
7&&
--&
--&&
--&
--\\
c ........ &
0.973&
12&&
0.983&
7&&
--&
--&&
--&
--\\
field ... &
&
3&&
&
4&&
&
15&&
&
7\\
\hline
\end{tabular}
}\label{tbl:confirmedsummary}
\begin{tabnote}
-- R.A. and DEC. are mean coordinates of the galaxies confirmed in the region.\\
-- $\langle z \rangle$ shows the median redshift of the galaxies confirmed. \\
-- N shows the number of member galaxies associated with the structures.\\
\end{tabnote}
\end{table*}

In total, we confirm 137 galaxies at $z$=0.8--1.1 among the 164
galaxies observed. Figure~\ref{fig:zspecVSzph} shows the comparison between
the photometric redshifts used in the target selection and the
spectroscopic redshifts determined by this study. The spectroscopic
redshifts tend to be systematically higher than the photometric
redshifts by $\sim$0.03, which indicates the importance of follow-up
spectroscopic observations.
The systematic difference between the photometric and spectroscopic
redshifts is similar for both red-sequence galaxies and star-forming
galaxies. However, the dispersion of the redshift difference is
slightly larger in the star-forming galaxies than the
red-sequence galaxies (lower panel of
figure~\ref{fig:zspecVSzph}). The larger dispersion in  
star-forming galaxies seems to be due to a larger fraction of
star-forming galaxies with photometric redshifts underestimated. 
This systematic offset of photometric redshift towards lower values
suggests that our photo-$z$ selection of galaxies can cause a lower
completeness of galaxies at lower redshifts among the redshift range
of 0.85--0.95. 
Indeed, we find that among the galaxies confirmed in
\citet{Lemaux2012}, 67\% of the galaxies at spectroscopic redshifts of
0.90--1.0 are selected by our photometric redshift selection, while
45\% of the galaxies at spectroscopic redshifts of 0.85--0.90 are selected.
This is why in figure~\ref{fig:map} the position of the
cluster-B at $z=0.865$ is not as overdense as the other clusters at
higher redshifts of $z\gtrsim0.9$ among the clusters presented
by \citet{Lemaux2012}. We make sure that the spatial distribution of
galaxies selected with the photometric redshift range shifted lower by
0.05 is more similar to the structures presented by
\citet{Lemaux2012}. 
The accuracy of the photometric redshifts demonstrates that
most of member galaxies associated with the galaxy clusters and groups at
$z\gtrsim0.9$ should be surely selected by the criteria we apply.  
 
Figure~\ref{fig:dis_zspec} shows redshift distributions of the
red-sequence galaxies (red histogram) and the star-forming galaxies
(blue histogram) confirmed in each overdensity region. The peaks of
redshift distribution are coincident between red-sequence galaxies and
star-forming galaxies. The individual redshift peaks show a wide range
of the fraction of red-sequence galaxies. Some overdensity regions
show about $\gtrsim$ 50\% of the fraction of red-sequence galaxies,
while some regions show as small as one-third. 
In particular, the S3 region shows the strong concentration of
star-forming galaxies,
suggesting that the cluster is less mature than the others. 
We note that there is a bias in the target selection where the
fraction of red-sequence galaxies observed in the S3 region is smaller
than those in the other regions (section
\ref{sec:MOSobsSubaruGemini}). Even if we correct for the bias, the
fraction of red-sequence galaxies in the S3 region is not as high as
those in the other regions.
The difference of the red fraction may reflect the structure formation
in this supercluster region, which is discussed in section~\ref{sec:discussions}
along with stellar population of red-sequence galaxies.

\subsection{Overview of the large-scale structure of CL1604 supercluster}

Figure~\ref{fig:map_summary} shows the close-up view of spatial
distribution, color-magnitude diagram, redshift distribution of the
confirmed galaxies, where the symbols are color-coded based on the
redshift range. The red-sequence galaxies in the N1 region show three
peaks in the distribution, suggesting three clusters/groups are
superposed and thus this region looks as if there is prominent
structure in the projected 2D map of galaxies (figures~\ref{fig:map}
and \ref{fig:map_summary}). The spatial distribution of galaxies with
different redshift ranges cannot be separated. The difference of the
redshifts in three peaks corresponds to more than 6000--7000 km
s$^{-1}$ away from each other. Since a typical velocity dispersion of
member galaxies in galaxy clusters is $\sim$ 1000--2000 km s$^{-1}$
depending on the cluster richness (e.g., \cite{Girardi1993}), the
structures at the different redshifts are likely to be independent.
The comoving distance between three peaks with different redshifts is
$\sim$ 100 comoving Mpc.    
The S1 region shows three small peaks. Interestingly, although these
are not as prominent as the peaks in the N1 region, the redshifts of
the peaks are quite similar. Given that both the structures are
located apart by $\sim$43 comoving Mpc from each other, there is 
likely to be three layers of the large-scale structures at three
different redshifts. The galaxies in the regions between the N1 and S1
also show similar redshift distribution
(figure~\ref{fig:specz_declination}, \cite{Lemaux2012},
\cite{Gal2008}).  
The south part of S2 that consists of galaxies at $z\approx0.9$ seems
to be connected to the structure of S3. The redshift distribution is
similar between the S2 and S3 overdense regions. 
Therefore, we confirm that this supercluster extends toward not only
the north-south direction but also the redshift direction, namely,
this is a complex three-dimensional structure more extended than
already-known. As shown in figure~\ref{fig:specz_declination}, since
there is a large spatial gap between galaxies observed by this study
and those confirmed by \citet{Lemaux2012}, it is difficult to conclude
whether the structures are sheets/filaments or the assembly of galaxy
clusters. However, since the local concentrations of galaxies are seen
in both of the declination and redshift directions, it must be certain
that the galaxy groups/clusters comprise the large scale structures.
These results may suggest that the assembly of galaxies
associated with the large-scale structures is synchronized over 50
comoving Mpc scale. The structures confirmed are summarized in
table~\ref{tbl:confirmedsummary}.        

\section{Formation history of the large-scale structure}
\label{sec:discussions}

\begin{figure*}
 \begin{center}
   \includegraphics[width=\linewidth]{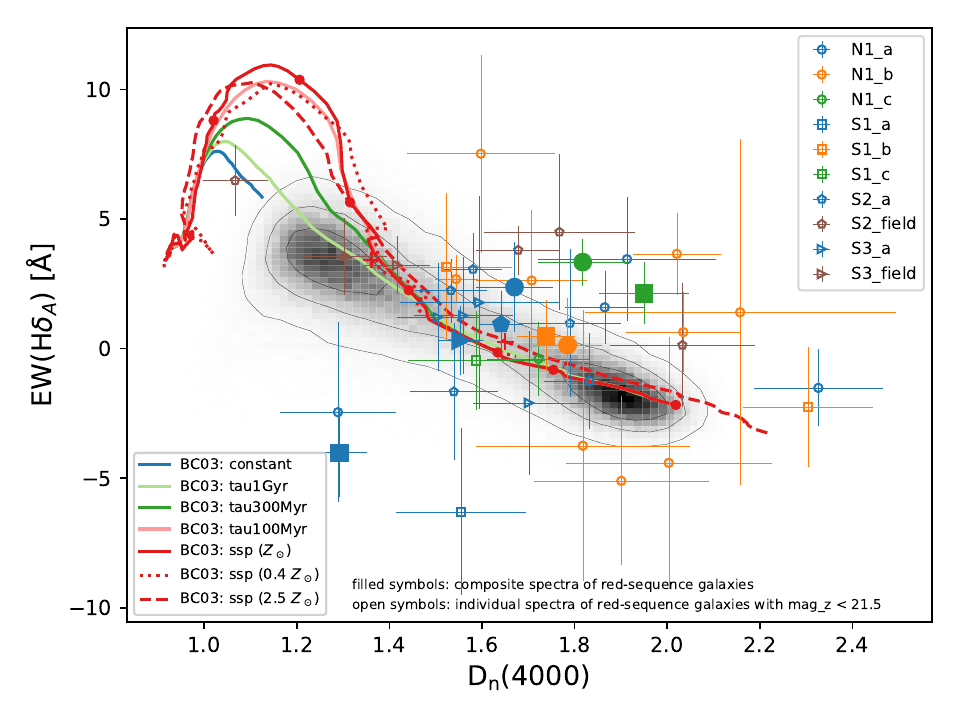}
 \end{center}
 \caption{Strength of Balmer H$\delta$ absorption line as a function
   of strength of 4000\AA\ break for red-sequence galaxies. 
   For individual galaxies, the only red-sequence galaxies with
   $z$-band mag brighter than 21.5 are investigated, while all of the
   red-sequence galaxies are used for the composite spectra. 
   We apply the definitions by \citet{Worthey1997} and
   \citet{Balogh1999} to measure the equivalent width of H$\delta$ and
   $D_n4000$, respectively. Symbol and color of the individual
   galaxies are different depending on the position and redshift
   (figure~\ref{fig:map_summary} and table~\ref{tbl:confirmedsummary}).
   The open symbols show the measurement for the individual galaxies,
   and the filled symbols show the measurement for the composite
   spectra of red-sequence galaxies (figure~\ref{fig:CompositeSpec}).
   The solid lines are model tracks of stellar population synthesis of
   \citet{bc03} with different star formation histories: simple
   stellar population (red), e-folding time of 0.1 Gyr (pink), 0.3 Gyr 
   (green), and 1.0 Gyr (yellow green) in exponentially declining star
   formation history, and constant star formation (blue). Stellar ages
   of 0.1, 0.5, 1, 2, 3, 5, and 10 Gyr are shown by red filled dots
   for SSP.   
   The dotted and dashed lines show model tracks of SSP with
   0.4$Z$\solar and 2.5$Z$\solar, respectively, where stellar age of
   2 Gyr is marked by the plus symbol.
   The gray-scale histogram shows the distribution of the SDSS galaxies
   at $z=$0.04--0.1 with the contours of 30, 68, and 90\% of the
   galaxies enclosed.
}\label{fig:D4Hd} 
\end{figure*}

\subsection{Stellar population of red-sequence galaxies}
\label{sec:StellarPopulation}

Stellar population of galaxies associated with the large-scale
structures are useful for understanding how the galaxies have evolved
along the hierarchical growth of host dark matter haloes. To do that,
we measure the strength of 4000\AA\ break of stellar continuum,
$D_n4000$, and strength of Balmer H$\delta$ absorption line.
The $D_n4000$ index is sensitive to stellar population age in the
sense that young galaxies have small 4000\AA\ break and old metal-rich
galaxies have large  4000\AA\ break \citep{Kauffmann2003b}.  
The equivalent width of H$\delta$, EW(H$\delta$), is sensitive to the
time scale of star formation history in the sense that the absorption
is the strongest 0.1--1.0 Gyr after burst of star formation
\citep{Kauffmann2003b}. Therefore, a diagram of H$\delta$ versus
$D_n4000$ is widely used to characterize stellar population of galaxies.

We apply the definition by \citet{Balogh1999} for the measurement of
$D_n4000$ index, where flux densities of stellar continuum at
3850--3950\AA\ and 4000--4100\AA\ are measured as a blue and red side
continuum level, respectively. There are several frequently used
definitions for EW(H$\delta$) depending on a difference in a
wavelength range of continuum level and/or absorption line \citep{Worthey1997,Fisher1998}. 
\citet{Worthey1997} define H$\delta_A$ as a measurement from the blue
continuum of 4041.60--4079.75 \AA, the red continuum of
4128.50--4161.00 \AA, and the absorption band of 4083.50--4122.25 \AA,
while \citet{Fisher1998} define it by the blue continuum of
4017.0--4057.0 \AA, the red continuum of 4153.0--4193.0 \AA, and the
absorption band of 4083.5--4122.25 \AA. The difference between the
two definitions of H$\delta$ is how the continuum level is
determined. We apply the definition by \citet{Worthey1997} for the
measurement of EW(H$\delta$) here. The positive value of EW means that
H$\delta$ is an absorption line. 

Figure~\ref{fig:D4Hd} shows the measurements of H$\delta$ spectral
feature as a function of those of $D_n4000$. 
The error shows the standard deviation of 100 measurements for the
individual spectrum with 1$\sigma$ noise added following a normal
distribution. We measure the spectral indices for only the 33
red-sequence galaxies with $z$-band mag brighter than 21.5, because
the other 22 fainter galaxies have too large uncertainty in the
measurement to determine the indices on an individual basis. 
The limiting magnitude corresponds to stellar mass of
$\gtrsim10^{10.8}$ M\solar, and 85\% of the bright galaxies have
stellar mass larger than $10^{11}$ M\solar.  
For comparison with the stellar population synthesis model, model
tracks of the \citet{bc03} spectra (updated version 2016\footnote{\url{http://www.bruzual.org/bc03/}}) 
with different star formation histories are also plotted in
figure~\ref{fig:D4Hd}. The star formation histories that we consider
are simple stellar population (SSP), exponentially declining one with 
e-folding time of 0.1 Gyr, 0.3 Gyr, and 1.0 Gyr, and constant star
formation. When the $D_n4000$ index ranges with the values less than
1.4, which corresponds to age $\lesssim$ 2 Gyr, EW(H$\delta$) can take
different values depending on the time scale of star
formation. However, after $\sim2$ Gyr, the tracks with different star
formation histories converge. Furthermore, as a reference, the
distribution of the line indices of galaxies at $z$=0.04--0.1, which
are extracted from the SDSS DR7 catalog of spectrum measurements
released by the
MPA-JHU\footnote{\url{https://wwwmpa.mpa-garching.mpg.de/SDSS/DR7/}}
is also shown \citep{Kauffmann2003b}. 

Most of the bright red-sequence galaxies have old ages, although the
individual galaxies show large variations on the
H$\delta$--$D_n4000$ diagram (figure~\ref{fig:D4Hd}). Different
symbols denote the different locations on the sky, and different colors
reflect different redshifts (the color coding is the same as in
figure~\ref{fig:map_summary}). The red-sequence galaxies in N1 region
tend to have larger values of $D_n4000$, compared with the galaxies in
southern structures. On the other hand, irrespective of the clusters
where the galaxies are located, there seems to be weak trend that the
red-sequence galaxies associated with the structure at similar
redshift (i.e., symbols with the same color) have similar spectral
indices.      
However, the individual measurements of EW(H$\delta$) and $D_n4000$
have large uncertainties even for bright galaxies with $z-$band mag
$<21.5$.  
Hereafter, we use composite spectra of the red-sequence galaxies in each
sub-structure to investigate the average properties. Then, we discuss
formation history of the CL1604 supercluster and stellar populations
of the red-sequence galaxies.  

\begin{figure*}
 \begin{center}
   \includegraphics[width=0.49\linewidth]{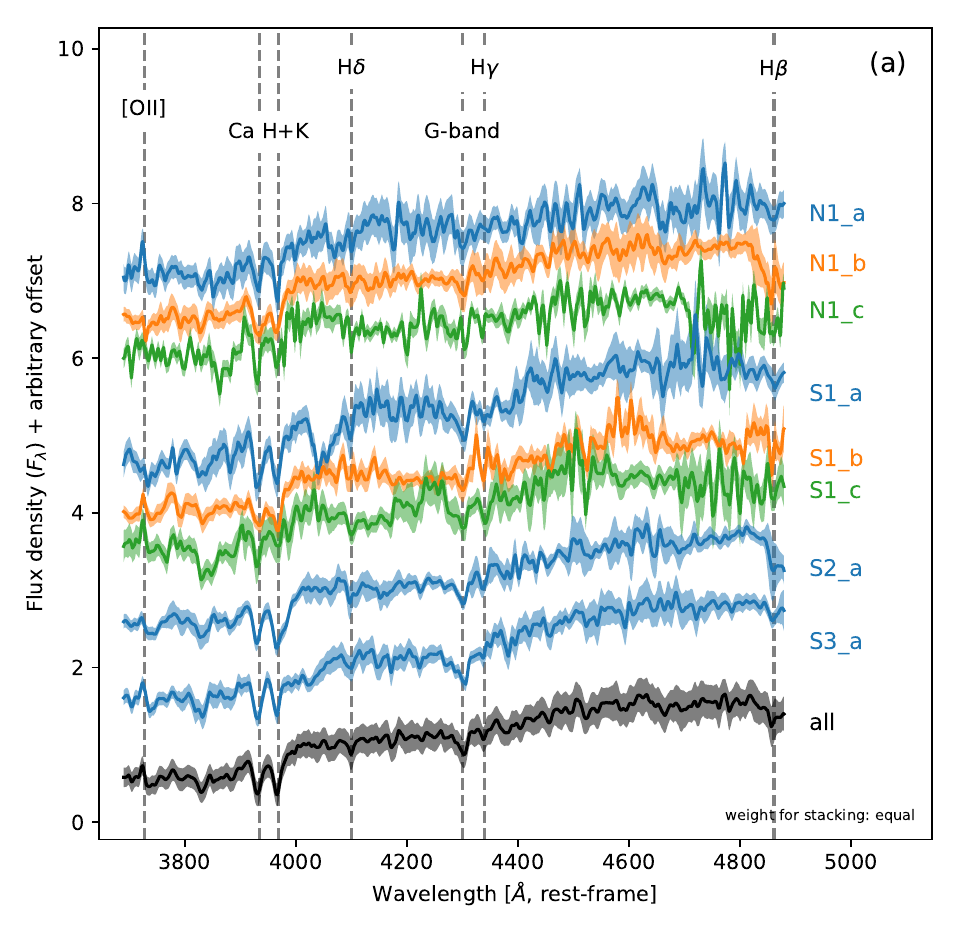}
   \includegraphics[width=0.49\linewidth]{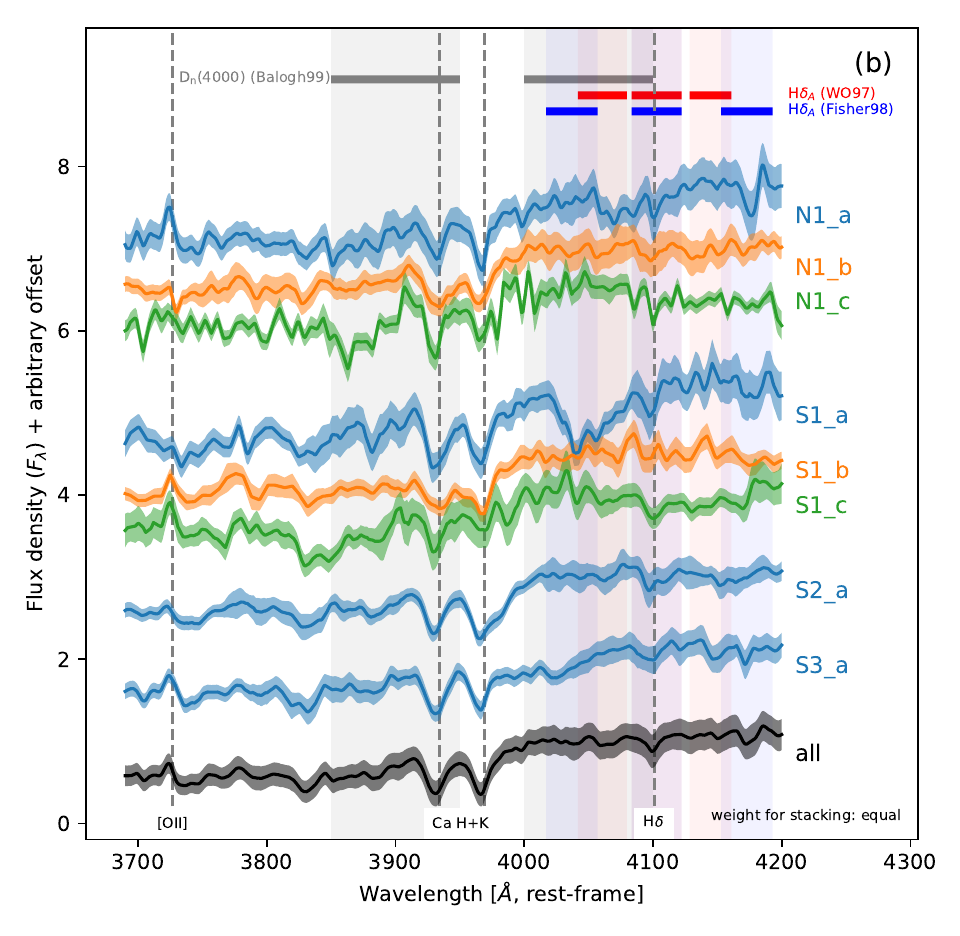}
 \end{center}
 \caption{The composite spectra of the red-sequence galaxies in each
   sub-structure (table~\ref{tbl:confirmedsummary}). The flux
   densities of each spectrum are normalized at 4000\AA, and then 
   stacked with an equal weight, i.e. we do not use a weight of
   1/$\sigma^2$ for stacking, because we prefer not to be biased to
   bright targets. The pale region around the spectrum shows the
   1$\sigma$ uncertainty.
   The color-coding in the spectra is the same as in figure~\ref{fig:map_summary}. In
   this plot, the spectra are shifted arbitrarily towards the vertical
   direction to avoid overlapping each other. The wavelength shown in
   the horizontal axis is in rest frame. The left figure (a) shows the
   spectra at the whole wavelength range of 3670--4900 \AA, while the
   right figure (b) shows the spectra at the limited range of 3670--4200
   \AA\ that are important for the spectral indices. In the figure
   (b), the bars and pale color regions show the wavelength ranges for
   each spectral index. 
   }\label{fig:CompositeSpec}   
\end{figure*}

To make composite spectra of the red-sequence galaxies in each
sub-structure, we first convert the wavelength in observed frame to 
that in rest frame and then normalize the flux densities of each
spectrum at 4000\AA. After that, we stack the spectra with an equal
weight. We do not use a weight of 1/$\sigma^2$ for stacking, because
we prefer not to be biased to bright targets. However, even if the
spectra are stacked with a weight of 1/$\sigma^2$, 
our discussions in this paper do not change. For this analysis, all of
the red-sequence galaxies confirmed are used, namely, including not
only bright galaxies but also fainter galaxies with $z-$band mag $>21.5$. 
The composite spectra of red-sequence galaxies are shown in
figure~\ref{fig:CompositeSpec}, and the number of galaxies stacked for
each composite spectrum is shown in table~\ref{tbl:specindicesstack}. 
The spectral indices of the composite spectra are measured in the same
manner as for the spectra of individual galaxies.
The strength of H$\delta$ for the composite spectra is also shown as a
function of that of 4000\AA\ break in figure~\ref{fig:D4Hd}, while the
measurements are listed in table~\ref{tbl:specindicesstack}.    
In the next section, we discuss the evolution of the large-scale
structure around the CL1604 supercluster.

\subsection{Implication for the evolution of the large-scale structure}
The composite spectra of red-sequence galaxies in the structures at
similar redshifts have comparable values of the spectral indices within
the errors. All but one (S1\_a) indicate the spectra with stellar population
older than 2 Gyr, which is comparable with the measurement by
\citet{Lemaux2012} for the already known CL1604 clusters. 
Although the red-sequence galaxies are located in the large-scale
structures over 50 Mpc scale, our results suggest that the formation
epoch is likely to be similar over the whole of structures.

There are several composite spectra of red-sequence galaxies with
spectral indices deviating from the model tracks towards the upper
right direction. Since the Balmer H$\delta$ is sensitive to the time
scale of star formation, one possibility is that the star-formation
activity is dependent on the sub-structures along the growth of
the large-scale structure. Figure~\ref{fig:dis_zspec} shows that there
is a large variation in the fraction of star-forming galaxies between
the substructures. The sub-structures at higher redshift, i.e., the
N1\_c and S1\_c, with the larger deviation may have higher fraction of
star-forming galaxies, although the S3\_a  with the spectral
indices on the model tracks also seems to be dominated by star-forming
galaxies. 
We should note that the target selection for our spectroscopic
observation can cause a bias in the fraction of the population
(see also the discussion of target selection bias in section
\ref{sec:redshift_measurement}). However, it is possible 
that the star formation activity, i.e., the fraction of star-forming
galaxies, is different between clusters depending on the assembly
process such as timing of accretion into more massive halo and
interaction with surroundings. 
In the model of stellar population synthesis, we assume the SSP or
smooth history of star formation declining exponentially with time,
which means that the model can not take account of the small starburst
occurred recently and/or on-going weak dusty star formation. 
Recent studies of post-starburst galaxies with strong Balmer
absorption which are selected by principal component analysis (PCA)
show that there are post-starburst galaxies in red sequence
rejuvenated by a minor merger with a gas-rich galaxy and dusty
starburst galaxies are a population of major contaminants of
post-starburst galaxies \citep{Pawlik2018,Pawlik2019}. Given the
facts, the second small starburst in the later phase can change the
strength of spectral indices \citep{Kauffmann2003b,Marcillac2006,Lemaux2012}.  
However, although the 4000\AA\ break (H$\delta$ absorption) gets small (strong)
soon after the burst, as time passes, the effect of the second
starburst becomes negligible. The spectral indices observed in the
continuum spectra cannot be explained by only stellar continuum with
multiple starbursts.  

If star-formation occurs in the galaxies, emission lines from  H{\sc ii} 
regions should be superposed on the stellar continuum spectra. In
fact, there are composite spectra showing [O{\sc ii}] emission line
(figure~\ref{fig:CompositeSpec}), suggesting that non-negligible star
formation activity. That is, the sample of the red-sequence galaxies
can include a large fraction of so-called `E+A' galaxies \citep{Dressler1999,Poggianti1999}.
The PCA for galaxy spectra shows that there are post-starburst
galaxies with emission lines \citep{Wild2007,Pawlik2018}. 
The red-sequence galaxies do not seem to be purely quiescent galaxies.
There is another possibility that the emission lines comes from the
existence of active galactic nuclei \citep{Yan2006}. They all suggest
that the star formation history of the red-sequence galaxies is not as 
simple as the assumption in the stellar population synthesis model,
but the red-sequence galaxies have more complex, various histories of
star formation.  

Both the 4000\AA\ break and H$\delta$ absorption are sensitive to
metallicity and dust attenuation as well \citep{Kauffmann2003b,Marcillac2006},
in the sense that more metal-rich and/or heavier dust attenuation make
the galaxies move slightly towards the upper right direction on the
H$\delta$--$D_n4000$ diagram. It is probable that dusty starburst
galaxies at higher redshifts make a larger contribution to the
spectral indices. Several previous studies reports that the dusty
starburst galaxies are found in the overdense regions at high redshifts \citep{Geach2009,Koyama2010,Koyama2011}.

\citet{Lemaux2012} found that the red-sequence galaxies in the central
clusters of this large-scale structure have stellar population
consistent with that by a single starburst. Therefore, by combining
the results with ours, a picture of the formation of the large-scale
structure is that the galaxies in the central massive halo are already
mature and the galaxies in the course of accretion into the denser
region can have a chance of rejuvenation with small multiple starburst
showing a complex star formation history.   

\begin{table}
\tbl{The spectral indices of the composite spectra.}{%
\begin{tabular}{ccccc}
\hline
cluster & N\footnotemark[$*$] & EW(H$\delta$)[$\AA$] & EW(H$\delta$)[$\AA$] & $D_n4000$ \\
& & WO97 & Fisher98 & Balogh99 \\
\hline
HSC-N1\_a& 11&  2.37$\pm$1.74&  3.43$\pm$1.78&  1.67$\pm$0.08 \\
HSC-N1\_b& 12&  0.13$\pm$1.82&  0.94$\pm$1.61&  1.78$\pm$0.07 \\
HSC-N1\_c& 2&  3.34$\pm$0.90&  2.80$\pm$0.87&  1.82$\pm$0.10 \\
HSC-S1\_a& 3& -4.03$\pm$1.71& -6.21$\pm$2.03&  1.29$\pm$0.06 \\
HSC-S1\_b& 3&  0.48$\pm$1.38& -2.75$\pm$1.32&  1.74$\pm$0.06 \\
HSC-S1\_c& 3&  2.14$\pm$1.19&  5.77$\pm$1.08&  1.95$\pm$0.10 \\
HSC-S2\_a& 7&  0.93$\pm$1.28&  0.69$\pm$1.37&  1.64$\pm$0.05 \\
HSC-S3\_a& 6&  0.31$\pm$1.32& -3.22$\pm$1.42&  1.55$\pm$0.05 \\
\hline
\end{tabular}
}\label{tbl:specindicesstack}
\begin{tabnote}
\footnotemark[$*$] The number of the red-sequence galaxies stacked.\\
\end{tabnote}
\end{table}

\section{Conclusions}
\label{sec:conclusions}

In this paper, we reveal the whole picture of the CL1604 supercluster at
$z\sim0.9$. Although the CL1604 supercluster was already known as a 
large-scale structure over $\sim26$ comoving Mpc consisting of three
galaxy clusters with $M_{cl}>10^{14} M_{\solar}$ and 
at least five galaxy groups with $M_{g}>10^{13} M_{\solar}$
\citep{Lemaux2012}, the HSC-SSP deep wide-field imaging survey reveals
the larger-scale structures extended to the north and south over more
than 50 comoving Mpc scale using galaxies selected with photometric
redshifts of $z \sim 0.9$. We then confirm that there are galaxy
clusters and/or groups in both the northern and southern side by
determining the redshifts of 55 red-sequence galaxies and 82
star-forming galaxies in total. It is found that the already-known 
CL1604 supercluster is a mere part of larger-scale structures and the
tip of the iceberg.  

We investigate stellar population of the red sequence galaxies by
measuring strength of 4000\AA\ break, $D_n4000$, and equivalent width
of Balmer H$\delta$ absorption line, EW(H$\delta$). Almost all of the
red-sequence galaxies brighter than 21.5 mag in $z$-band show the old
stellar population with $D_n4000>1.5$, which corresponds to stellar
age of $\gtrsim2$ Gyr. The comparison of composite
spectra of red-sequence galaxies in individual clusters show that even
if the red-sequence galaxies are located $>$50 Mpc apart, they at a
similar redshift have similar stellar population age. However, there
is a large variation in the star formation history. Therefore, it is
likely that galaxies associated with the large-scale structures at 50
Mpc scale formed at almost the same time, have assembled into the 
denser regions, and then evolved along the hierarchical growth of the
cosmic web.  

\begin{ack}
We thank the anonymous referee for providing constructive comments and
suggestions. We thank Dr. Brian C.~Lemaux and Prof. Lori M.~Lubin for
kindly providing us with the catalog of the confirmed galaxies used in
\citet{Lemaux2012} and \citet{Lemaux2018}. We are also grateful to
Dr. Brian C.~Lemaux for his comments and suggestions. 
All of them are useful to improve this paper.  

This paper is based on data collected at Subaru Telescope and
Gemini-North Telescope via the time exchange program between Subaru
and the Gemini Observatory and data retrieved from the Hyper
Suprime-Cam (HSC) data archive system, which is operated by the Subaru
Telescope and Astronomy Data Center at National Astronomical
Observatory of Japan. 

The HSC collaboration includes the astronomical 
communities of Japan and Taiwan, and Princeton University. The HSC
instrumentation and software were developed by the National
Astronomical Observatory of Japan (NAOJ), the Kavli Institute for the
Physics and Mathematics of the Universe (Kavli IPMU), the University
of Tokyo, the High Energy Accelerator Research Organization (KEK), the
Academia Sinica Institute for Astronomy and Astrophysics in Taiwan
(ASIAA), and Princeton University. Funding was contributed by the
FIRST program from Japanese Cabinet Office, the Ministry of Education,
Culture, Sports, Science and Technology (MEXT), the Japan Society for
the Promotion of Science (JSPS),  Japan Science and Technology Agency
(JST),  the Toray Science  Foundation, NAOJ, Kavli IPMU, KEK, ASIAA,
and Princeton University.  
The Pan-STARRS1 Surveys (PS1) have been made possible through
contributions of the Institute for Astronomy, the University of
Hawaii, the Pan-STARRS Project Office, the Max-Planck Society and its
participating institutes, the Max Planck Institute for Astronomy,
Heidelberg and the Max Planck Institute for Extraterrestrial Physics,
Garching, The Johns Hopkins University, Durham University, the
University of Edinburgh, Queen's University Belfast, the
Harvard-Smithsonian Center for Astrophysics, the Las Cumbres
Observatory Global Telescope Network Incorporated, the National
Central University of Taiwan, the Space Telescope Science Institute,
the National Aeronautics and Space Administration under Grant
No. NNX08AR22G issued through the Planetary Science Division of the
NASA Science Mission Directorate, the National Science Foundation
under Grant No. AST-1238877, the University of Maryland, and Eotvos
Lorand University (ELTE). 
This paper makes use of software developed for the Large Synoptic
Survey Telescope. We thank the LSST Project for making their code
available as free software at \url{http://dm.lsst.org}.
\end{ack}

\appendix 

\section{The measurement of H$\delta$ with the different spectral index}
\label{sec:appendix_SpecIndex}

\begin{figure}
 \begin{center}
   \includegraphics[width=\linewidth]{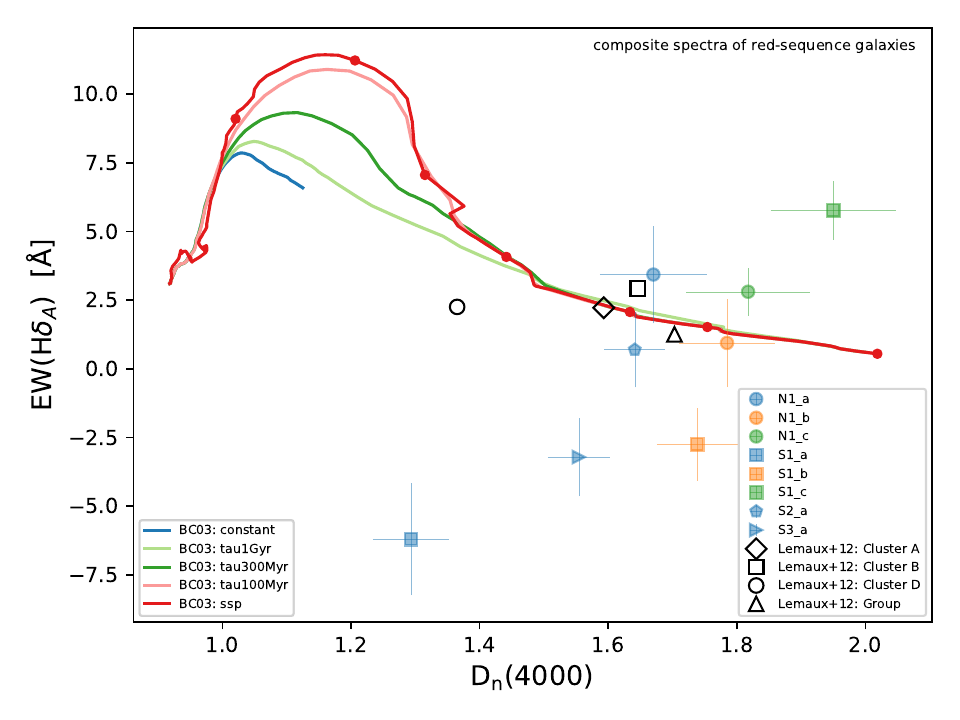}
 \end{center}
 \caption{Same as figure~\ref{fig:D4Hd}, but the definition by
   \citet{Fisher1998} is used for the measurement of Balmer H$\delta$
   absorption line. \citet{Lemaux2012} used the definition to measure
   the strength of H$\delta$ line. The open symbols shows the results
   of \citet{Lemaux2012}. 
 }\label{fig:D4Hd_L12}
\end{figure}

Here, we use the definition by \citet{Fisher1998} for the measurement
of Balmer H$\delta$ absorption line in the composite spectra. This is
because \citet{Lemaux2012} use the definition of H$\delta$ for the
galaxies in the already-known central structures. 
Compared with the measurement with the definition by
\citet{Worthey1997}, there is a large dispersion in the H$\delta$
strength (figure~\ref{fig:D4Hd_L12} and
table~\ref{tbl:specindicesstack}). As shown in
figure~\ref{fig:CompositeSpec}, the difference between the indices
is the wavelength range used to estimate stellar continuum 
level. Since the index of \citet{Fisher1998} use the wavelength range
apart from the H$\delta$ rather than the index of \citet{Worthey1997},
the measurement of H$\delta$ is more sensitive to the stellar continuum,
which could cause the larger dispersion. Although the reason of the
large dispersion is unknown, there are several sky lines around
H$\delta$ and the accuracy of subtraction of the sky lines can affect
the signal-to-noise ratio of the continuum spectrum. Note that we did
not use the nod and shuffle mode in the spectroscopic observations. 
Therefore, we use the definition by \citet{Worthey1997} for the
discussion in section~\ref{sec:discussions}. The measurements by
\citet{Lemaux2012} are plotted in figure~\ref{fig:D4Hd_L12} for
comparison with our results. 

\section{The catalogs of galaxies confirmed}
\label{sec:appendix_Catalog}

Tables \ref{tbl:RSGs} and \ref{tbl:SFGs} shows the catalogs of the
red-sequence galaxies and star-forming galaxies that are confirmed by
the observations, where the information of {\tt object\_id}, coordinates,
magnitude, and color are extracted from the HSC-SSP PDR2 data
\citep{HSCSSPDR2}. 
For reference, only {\tt object\_id} of the S16A data is shown.
The S16A data are internally released and used in the original target
selection described in section~\ref{sec:obsanddata}. 
For the red-sequence galaxies, the redshift, equivalent width of
H$\delta$, and $D_n4000$ index are measured from the spectra of
individual galaxies in subsections~\ref{sec:redshift_measurement} and
\ref{sec:StellarPopulation}. For the star-forming galaxies, the last
column of table~\ref{tbl:SFGs} shows the emission lines detected in
the spectra of individual galaxies. In the case that a single emission
line is detected, we assume that the line is [O\emissiontype{II}] and
then the redshift is measured. If required, additional information of
the individual galaxies can be extracted from the HSC-SSP PDR2
database\footnotemark[\ref{foot:hscsspwebsite}], based on {\tt object\_id}. 

\onecolumn
\footnotesize
\begin{landscape}
\begin{longtable}[c]{cccccccccccc}
\caption{A list of 55 red-sequence galaxies confirmed. The object ID in the first column is from the S16A release data and the other information is from the PDR2 data and spectrum.\\}\label{tbl:RSGs}\\
\hline
object\_id (S16A) & object\_id (PDR2) & R.A. & Dec. & cluster & z mag. & r-z & spectrograph & redshift & EW(H$\delta$)[$\AA$] & EW(H$\delta$)[$\AA$] & $D_n4000$ \\
& & & & & & & & & WO97 & Fisher98 & Balogh99 \\
\endfirsthead
\hline
\endhead
\hline
\endfoot
\hline
\endlastfoot
\hline
69617112015996160& 69617112015996479& 16:03:40.395& +42:38:38.425& S3& 21.02& 2.14&  GMOS& 0.890& -2.10$\pm$2.77& -6.05$\pm$2.78&  1.70$\pm$0.12 \\
69617112015996904& 69617112015998074& 16:03:33.596& +42:39:13.880& S3& 21.45& 2.26&  GMOS& 0.907&  1.77$\pm$4.10& -4.20$\pm$4.55&  1.59$\pm$0.17 \\
69617112015997316& 69617112015998454& 16:03:46.199& +42:39:30.601& S3& 20.20& 2.15&  GMOS& 0.890&  1.21$\pm$2.08& -4.10$\pm$2.65&  1.50$\pm$0.09 \\
69617112015972551& 69617112015998558& 16:03:40.219& +42:39:40.821& S3& 21.78& 2.08&  GMOS& 0.910& ---& ---& --- \\
69617112015997717& 69617112015998746& 16:03:51.500& +42:39:53.591& S3& 21.78& 2.04&  GMOS& 0.898& ---& ---& --- \\
69617112015997796& 69617112015998848& 16:03:47.229& +42:39:57.713& S3& 21.11& 2.03&  GMOS& 1.010&  3.21$\pm$1.13&  4.49$\pm$0.90&  1.42$\pm$0.07 \\
69617112015997981& 69617112015999018& 16:03:34.390& +42:40:07.680& S3& 21.00& 2.38&  GMOS& 0.903&  1.27$\pm$2.26&  1.05$\pm$2.21&  1.56$\pm$0.11 \\
69617116310940051& 69617116310939901& 16:03:55.668& +42:50:28.562& S2& 20.70& 1.98&  GMOS& 0.911&  3.06$\pm$1.38&  4.00$\pm$0.94&  1.58$\pm$0.06 \\
69617116310938517& 69617116310964806& 16:03:46.843& +42:48:53.180& S2& 21.94& 2.14&  GMOS& 0.918& ---& ---& --- \\
69617116310965204& 69617116310965707& 16:03:48.209& +42:49:05.619& S2& 20.35& 2.15&  GMOS& 0.920&  2.24$\pm$1.23&  2.90$\pm$1.26&  1.53$\pm$0.08 \\
69617116310965418& 69617116310966045& 16:03:47.753& +42:49:34.080& S2& 21.54& 2.14&  GMOS& 0.917& ---& ---& --- \\
69617116310965615& 69617116310966133& 16:03:46.631& +42:49:45.052& S2& 21.80& 1.84&  GMOS& 0.919& ---& ---& --- \\
69617116310965997& 69617116310966570& 16:03:46.099& +42:50:06.593& S2& 21.47& 2.10&  GMOS& 0.916&  0.98$\pm$2.86& -0.04$\pm$2.70&  1.79$\pm$0.11 \\
69617116310967412& 69617116310967357& 16:03:37.715& +42:51:07.775& S2& 21.04& 1.92&  GMOS& 0.988&  6.49$\pm$1.36&  7.25$\pm$1.32&  1.07$\pm$0.07 \\
69617116310966811& 69617116310967364& 16:03:37.056& +42:50:48.871& S2& 21.28& 2.09&  GMOS& 0.918& -1.66$\pm$2.64& -2.71$\pm$3.12&  1.54$\pm$0.09 \\
69617116310966802& 69617116310967498& 16:03:50.289& +42:50:48.359& S2& 20.45& 2.38&  GMOS& 0.988&  3.80$\pm$0.95&  4.37$\pm$1.28&  1.68$\pm$0.09 \\
69617120605920352& 69617120605921712& 16:03:42.419& +42:51:59.018& S2& 21.27& 1.94&  GMOS& 0.989&  3.55$\pm$1.47&  8.34$\pm$1.53&  1.30$\pm$0.09 \\
69617120605921126& 69617120605922446& 16:03:40.346& +42:52:36.029& S2& 21.46& 2.05&  GMOS& 0.944&  0.13$\pm$2.41&  2.98$\pm$2.15&  2.03$\pm$0.16 \\
69617120605921738& 69617120605923049& 16:03:35.961& +42:53:02.321& S2& 21.27& 2.56&  GMOS& 0.948&  4.50$\pm$3.00&  2.82$\pm$2.96&  1.77$\pm$0.16 \\
69617120605922187& 69617120605923406& 16:03:16.712& +42:53:14.113& S1& 20.75& 2.15&  GMOS& 0.985& -0.46$\pm$1.93&  3.32$\pm$2.12&  1.59$\pm$0.15 \\
69617120605922445& 69617120605923568& 16:03:14.618& +42:53:28.231& S1& 22.09& 1.91&  GMOS& 0.983& ---& ---& --- \\
69617120605923272& 69617120605924511& 16:03:24.090& +42:54:13.470& S1& 21.91& 2.13&  GMOS& 0.921& ---& ---& --- \\
69617120605925007& 69617120605926484& 16:03:14.817& +42:55:39.814& S1& 22.15& 1.93&  GMOS& 0.887& ---& ---& --- \\
69617120605925117& 69617120605926513& 16:03:18.248& +42:55:42.545& S1& 21.81& 1.77&  GMOS& 0.842& ---& ---& --- \\
69617120605925431& 69617120605926983& 16:03:20.049& +42:56:05.017& S1& 21.00& 2.13&  GMOS& 0.949& -2.26$\pm$2.30& -1.63$\pm$2.59&  2.30$\pm$0.14 \\
69617120605926287& 69617120605927999& 16:03:10.009& +42:56:42.710& S1& 21.39& 2.06&  GMOS& 0.883& -6.30$\pm$3.22& -9.04$\pm$3.96&  1.55$\pm$0.14 \\
69617258044872416& 69617258044871201& 16:03:09.522& +42:52:03.196& S1& 20.86& 2.10&  GMOS& 0.930&  3.15$\pm$2.86& -0.53$\pm$2.83&  1.52$\pm$0.14 \\
69617258044872673& 69617258044871693& 16:03:08.446& +42:52:16.457& S1& 21.95& 1.97&  GMOS& 0.975& ---& ---& --- \\
69617258044877131& 69617258044875360& 16:03:09.587& +42:55:22.273& S1& 21.74& 2.00&  GMOS& 0.883& ---& ---& --- \\
70399543388171802& 70399543388172272& 16:03:52.790& +43:40:02.320& N1& 21.09& 1.96&  GMOS& 0.887& -2.46$\pm$3.47& -8.67$\pm$3.60&  1.29$\pm$0.13 \\
70399543388149323& 70399543388173819& 16:03:53.691& +43:41:08.039& N1& 21.76& 2.09&  GMOS& 0.933& ---& ---& --- \\
70399543388174556& 70399543388174278& 16:03:47.435& +43:41:33.969& N1& 20.59& 2.18& FOCAS& 0.929&  0.64$\pm$1.55&  1.42$\pm$1.50&  2.03$\pm$0.12 \\
70399543388175946& 70399543388175701& 16:03:52.016& +43:42:35.393& N1& 21.48& 2.09& FOCAS& 0.943& -5.10$\pm$3.26& -2.72$\pm$2.79&  1.90$\pm$0.19 \\
70399543388176326& 70399543388176068& 16:03:50.260& +43:42:54.783& N1& 21.85& 2.02&  GMOS& 0.871& ---& ---& --- \\
70399543388176450& 70399543388176321& 16:03:51.211& +43:43:07.084& N1& 20.51& 2.12& FOCAS& 0.874& -1.52$\pm$1.48& -0.32$\pm$1.18&  2.33$\pm$0.14 \\
70399543388176621& 70399543388176379& 16:03:56.928& +43:43:12.446& N1& 20.72& 2.21&  GMOS& 0.930&  2.62$\pm$2.72& -1.48$\pm$2.92&  1.71$\pm$0.12 \\
70399543388177131& 70399543388176965& 16:04:02.224& +43:43:42.694& N1& 21.06& 2.17&  GMOS& 0.928& -3.75$\pm$5.22& -1.37$\pm$3.99&  1.82$\pm$0.23 \\
70399543388177394& 70399543388177274& 16:03:51.068& +43:43:55.430& N1& 21.43& 1.93& FOCAS& 0.879&  3.45$\pm$2.40&  3.19$\pm$2.47&  1.91$\pm$0.19 \\
70399543388177404& 70399543388177291& 16:03:55.002& +43:43:56.749& N1& 21.62& 2.22&  GMOS& 0.921& ---& ---& --- \\
70399543388177612& 70399543388177324& 16:04:00.444& +43:44:08.426& N1& 22.32& 2.07& FOCAS& 0.975& ---& ---& --- \\
70399543388177604& 70399543388177513& 16:03:55.311& +43:44:03.279& N1& 21.48& 2.15& FOCAS& 0.912&  7.52$\pm$3.82&  9.16$\pm$3.47&  1.60$\pm$0.16 \\
70399543388177641& 70399543388177524& 16:03:54.306& +43:44:04.750& N1& 21.37& 2.25&  GMOS& 0.928& -4.41$\pm$4.87&  0.19$\pm$3.43&  2.00$\pm$0.22 \\
70399543388177700& 70399543388177564& 16:03:52.119& +43:44:17.076& N1& 20.54& 1.88& FOCAS& 0.881& -1.26$\pm$1.85&  1.98$\pm$2.04&  1.83$\pm$0.10 \\
70399543388177702& 70399543388177566& 16:03:52.277& +43:44:18.158& N1& 20.06& 2.71& FOCAS& 0.884&  1.59$\pm$1.40& -0.85$\pm$1.49&  1.86$\pm$0.09 \\
70399543388177737& 70399543388177865& 16:03:54.781& +43:44:29.036& N1& 21.66& 2.07& FOCAS& 0.888& ---& ---& --- \\
70399543388152808& 70399543388177938& 16:03:48.655& +43:44:24.062& N1& 21.99& 2.16&  GMOS& 0.929& ---& ---& --- \\
70399543388178162& 70399543388178065& 16:04:02.859& +43:44:25.165& N1& 22.21& 2.15& FOCAS& 0.881& ---& ---& --- \\
70399543388178592& 70399543388178703& 16:03:49.598& +43:45:00.030& N1& 21.41& 2.02&  GMOS& 0.930&  1.40$\pm$6.67& -5.01$\pm$6.71&  2.16$\pm$0.34 \\
70399543388178729& 70399543388178779& 16:03:44.668& +43:44:58.905& N1& 20.44& 2.20&  GMOS& 0.970& -0.40$\pm$1.43&  1.85$\pm$1.45&  1.72$\pm$0.11 \\
70399543388154281& 70399543388179847& 16:03:53.212& +43:45:41.672& N1& 21.92& 2.23&  GMOS& 0.867& ---& ---& --- \\
70405040946315068& 70405040946315286& 16:04:09.272& +43:43:43.364& N1& 22.14& 2.17& FOCAS& 0.882& ---& ---& --- \\
70405040946315628& 70405040946315935& 16:04:04.083& +43:44:17.812& N1& 20.11& 2.14& FOCAS& 0.920&  3.65$\pm$1.58&  3.24$\pm$1.36&  2.02$\pm$0.09 \\
70405040946316160& 70405040946316384& 16:04:04.581& +43:44:30.831& N1& 22.07& 2.26& FOCAS& 0.854& ---& ---& --- \\
70405040946316183& 70405040946316459& 16:04:08.803& +43:44:36.930& N1& 20.94& 1.93& FOCAS& 0.920&  2.68$\pm$0.93&  2.22$\pm$0.93&  1.54$\pm$0.05 \\
70405040946317856& 70405040946318093& 16:04:05.887& +43:45:44.182& N1& 21.76& 2.06& FOCAS& 0.824& ---& ---& --- \\
\end{longtable}
\end{landscape}

\begin{landscape}
\begin{longtable}[c]{cccccccccccc}
\caption{The same as table \ref{tbl:RSGs}, but for 82 star-forming galaxies confirmed. In the last column, detection or non-detection of the line is shown by $\circ$ or $\times$, respectively, and --- means that the spectrum does not cover the wavelength for the emission line.\\}\label{tbl:SFGs}\\
\hline
object\_id (S16A) & object\_id (PDR2) & R.A. & Dec. & cluster & z mag. & r-z & spectrograph & redshift & \multicolumn{3}{c}{emission lines detected$^*$} \\
\cline{10-12}
 & &  &  &  &  &  &  &  & [O\emissiontype{II}]3727 & H$\beta$ & [O\emissiontype{III}]5007 \\
\endfirsthead
\hline
\endhead
\hline
\endfoot
\hline
\endlastfoot
\hline
69617112015971747& 69617112015972223& 16:03:41.447& +42:38:53.698& S3& 22.20& 1.01&  GMOS& 0.891& $\circ$& $\circ$& $\circ$ \\
69617112015998043& 69617112015973587& 16:03:43.742& +42:40:12.950& S3& 23.89& 1.13&  GMOS& 0.878& $\circ$& $\circ$& $\circ$ \\
69617112015994346& 69617112015995528& 16:03:41.877& +42:37:15.491& S3& 22.85& 0.84&  GMOS& 0.895& $\circ$& $\circ$& $\circ$ \\
69617112015995059& 69617112015996295& 16:03:48.869& +42:37:57.262& S3& 22.22& 0.47&  GMOS& 0.895& $\circ$& $\circ$& $\circ$ \\
69617112015995799& 69617112015996841& 16:03:34.176& +42:38:16.865& S3& 22.36& 1.11&  GMOS& 0.902& $\circ$& $\circ$& $\times$ \\
69617112015995802& 69617112015996843& 16:03:35.001& +42:38:20.476& S3& 22.20& 1.00&  GMOS& 1.009& $\circ$& $\circ$& $\times$ \\
69617112015995909& 69617112015996963& 16:03:35.844& +42:38:25.340& S3& 23.41& 0.91&  GMOS& 0.986& $\circ$& $\times$& $\times$ \\
69617112015996131& 69617112015997201& 16:03:35.224& +42:38:37.830& S3& 21.14& 1.15&  GMOS& 0.884& $\circ$& $\circ$& $\circ$ \\
69617112015997302& 69617112015998278& 16:03:43.310& +42:39:31.621& S3& 21.83& 1.04&  GMOS& 0.877& $\circ$& $\circ$& $\circ$ \\
69617112015997635& 69617112015998643& 16:03:55.195& +42:39:47.280& S3& 22.51& 0.96&  GMOS& 0.916& $\circ$& $\times$& $\circ$ \\
69617112015997683& 69617112015998696& 16:03:33.160& +42:39:53.549& S3& 22.81& 0.88&  GMOS& 0.903& $\circ$& $\times$& $\times$ \\
69617112015997789& 69617112015998795& 16:03:56.552& +42:39:55.355& S3& 22.88& 0.81&  GMOS& 0.932& $\circ$& $\times$& $\circ$ \\
69617112015997875& 69617112015998896& 16:03:48.370& +42:40:02.643& S3& 23.27& 0.63&  GMOS& 0.907& $\circ$& $\circ$& $\circ$ \\
69617112015998121& 69617112015999163& 16:03:30.843& +42:40:17.391& S3& 23.13& 1.09&  GMOS& 0.902& $\circ$& $\times$& --- \\
69617112015998126& 69617112015999228& 16:03:37.178& +42:40:20.260& S3& 23.22& 1.07&  GMOS& 0.854& $\circ$& $\times$& $\circ$ \\
69617116310940018& 69617116310939871& 16:03:33.639& +42:50:26.516& S2& 22.84& 1.43&  GMOS& 0.881& $\circ$& $\circ$& $\circ$ \\
69617116310940107& 69617116310939957& 16:03:36.047& +42:50:31.268& S2& 23.60& 1.01&  GMOS& 0.847& $\circ$& $\times$& $\times$ \\
69617116310954407& 69617116310954387& 16:03:37.944& +42:41:30.252& S3& 23.22& 1.15&  GMOS& 0.901& $\circ$& $\circ$& $\times$ \\
69617116310955103& 69617116310955133& 16:03:37.582& +42:42:06.415& S3& 23.82& 0.71&  GMOS& 0.903& $\circ$& $\circ$& --- \\
69617116310955318& 69617116310955367& 16:03:37.297& +42:42:16.047& S3& 22.12& 0.98&  GMOS& 0.901& $\circ$& $\circ$& --- \\
69617116310955586& 69617116310955661& 16:03:41.623& +42:42:39.416& S3& 22.18& 0.92&  GMOS& 0.911& $\circ$& $\circ$& $\times$ \\
69617116310955800& 69617116310955864& 16:03:36.450& +42:42:44.562& S3& 23.17& 1.05&  GMOS& 1.032& $\circ$& ---& --- \\
69617116310956200& 69617116310956331& 16:03:43.304& +42:42:58.427& S3& 22.35& 0.96&  GMOS& 0.949& $\circ$& $\times$& --- \\
69617116310965358& 69617116310965934& 16:03:44.390& +42:49:27.582& S2& 22.47& 0.79&  GMOS& 0.897& $\circ$& $\circ$& $\circ$ \\
69617116310965750& 69617116310966450& 16:03:46.985& +42:49:58.922& S2& 23.07& 1.20&  GMOS& 0.915& $\circ$& $\times$& $\times$ \\
69617116310965831& 69617116310966562& 16:03:38.822& +42:49:57.133& S2& 23.33& 0.42&  GMOS& 0.883& $\circ$& $\circ$& $\circ$ \\
69617116310966485& 69617116310967306& 16:03:55.025& +42:50:33.612& S2& 22.23& 1.33&  GMOS& 0.910& $\circ$& $\circ$& --- \\
69617116310940789& 69617116310968260& 16:03:46.155& +42:51:12.847& S2& 23.72& 1.20&  GMOS& 0.944& $\circ$& $\times$& $\times$ \\
69617116310967672& 69617116310968440& 16:03:54.928& +42:51:19.114& S2& 23.72& 0.89&  GMOS& 0.957& $\circ$& ---& --- \\
69617116310967867& 69617116310968531& 16:03:44.318& +42:51:29.132& S2& 23.38& 1.11&  GMOS& 0.956& $\circ$& $\times$& $\circ$ \\
69617120605898714& 69617120605898781& 16:03:54.415& +42:53:31.323& S2& 22.37& 0.86&  GMOS& 1.054& $\circ$& ---& --- \\
69617120605899000& 69617120605899079& 16:03:17.325& +42:53:51.542& S1& 22.10& 1.25&  GMOS& 0.983& $\circ$& $\circ$& $\times$ \\
69617120605899227& 69617120605899310& 16:03:22.279& +42:54:05.236& S1& 22.93& 0.87&  GMOS& 0.949& $\circ$& $\times$& $\circ$ \\
69617120605899657& 69617120605899755& 16:03:30.230& +42:54:31.474& S1& 22.48& 0.52&  GMOS& 0.891& $\circ$& $\circ$& $\circ$ \\
69617120605899803& 69617120605899905& 16:03:17.542& +42:54:39.815& S1& 23.59& 0.97&  GMOS& 0.987& $\circ$& $\times$& $\times$ \\
69617120605897004& 69617120605921438& 16:03:11.168& +42:51:51.535& S1& 22.68& 1.00&  GMOS& 0.979& $\circ$& $\times$& $\times$ \\
69617120605920797& 69617120605922105& 16:03:27.971& +42:52:16.585& S1& 22.80& 1.19&  GMOS& 0.946& $\circ$& $\times$& $\times$ \\
69617120605920815& 69617120605922137& 16:03:47.543& +42:52:16.158& S2& 23.53& 0.42&  GMOS& 0.954& $\circ$& $\times$& $\circ$ \\
69617120605921069& 69617120605922379& 16:03:28.520& +42:52:31.786& S1& 22.87& 0.94&  GMOS& 1.033& $\circ$& $\circ$& $\circ$ \\
69617120605921347& 69617120605922626& 16:03:47.224& +42:52:49.718& S2& 23.09& 1.10&  GMOS& 1.071& $\circ$& $\times$& --- \\
69617120605921273& 69617120605922639& 16:03:31.879& +42:52:46.904& S2& 22.65& 0.82&  GMOS& 0.989& $\circ$& $\circ$& $\times$ \\
69617120605921450& 69617120605922659& 16:03:15.737& +42:52:59.612& S1& 23.06& 0.99&  GMOS& 0.882& $\circ$& $\circ$& $\circ$ \\
69617120605922162& 69617120605923361& 16:03:56.630& +42:53:12.514& S2& 21.81& 1.07&  GMOS& 0.956& $\circ$& ---& --- \\
69617120605922575& 69617120605923738& 16:03:55.681& +42:53:24.606& S2& 21.63& 1.58&  GMOS& 0.987& $\circ$& ---& --- \\
69617120605922682& 69617120605923835& 16:03:16.698& +42:53:38.852& S1& 22.68& 1.02&  GMOS& 0.885& $\circ$& $\circ$& $\times$ \\
69617120605922760& 69617120605923979& 16:03:49.229& +42:53:40.665& S2& 22.78& 0.89&  GMOS& 0.899& $\circ$& $\circ$& $\circ$ \\
69617120605922974& 69617120605924154& 16:03:12.469& +42:53:53.035& S1& 21.38& 1.16&  GMOS& 0.846& $\circ$& $\circ$& $\circ$ \\
69617120605923017& 69617120605924233& 16:03:49.656& +42:53:57.493& S2& 21.75& 0.73&  GMOS& 0.918& $\circ$& $\circ$& $\circ$ \\
69617120605924331& 69617120605925643& 16:03:14.368& +42:55:04.030& S1& 23.13& 0.65&  GMOS& 0.888& $\circ$& $\circ$& $\circ$ \\
69617120605924664& 69617120605926078& 16:03:19.527& +42:55:24.706& S1& 22.37& 1.04&  GMOS& 1.024& $\circ$& $\circ$& $\times$ \\
69617120605925260& 69617120605926785& 16:03:20.346& +42:55:51.081& S1& 23.35& 0.94&  GMOS& 0.950& $\circ$& $\times$& $\times$ \\
69617120605926252& 69617120605927966& 16:03:15.560& +42:56:41.023& S1& 22.42& 0.84&  GMOS& 0.921& $\circ$& $\times$& $\circ$ \\
69617120605926493& 69617120605928199& 16:03:14.152& +42:56:53.532& S1& 22.74& 1.22&  GMOS& 0.985& $\circ$& $\times$& $\times$ \\
69617258044851644& 69617258044851490& 16:03:02.811& +42:53:06.735& S1& 22.00& 0.86&  GMOS& 0.870& $\circ$& $\circ$& --- \\
70399543388150085& 70399543388150126& 16:03:50.626& +43:41:51.193& N1& 22.91& 1.03& FOCAS& 0.925& $\circ$& $\times$& $\circ$ \\
70399543388150702& 70399543388150768& 16:03:47.106& +43:42:28.087& N1& 21.89& 1.09& FOCAS& 0.900& $\circ$& $\circ$& $\times$ \\
70399543388152173& 70399543388152246& 16:04:02.804& +43:43:47.343& N1& 23.63& 0.62& FOCAS& 0.922& $\circ$& $\times$& $\circ$ \\
70399543388152733& 70399543388152811& 16:03:58.077& +43:44:19.520& N1& 21.39& 1.38& FOCAS& 0.878& $\circ$& $\circ$& $\times$ \\
70399543388171595& 70399543388172114& 16:03:47.840& +43:39:48.928& N1& 22.37& 1.05&  GMOS& 0.886& $\circ$& $\times$& $\times$ \\
70399543388172392& 70399543388172832& 16:03:51.969& +43:40:27.033& N1& 21.85& 0.89&  GMOS& 1.003& $\circ$& $\circ$& $\circ$ \\
70399543388173927& 70399543388173733& 16:03:44.656& +43:41:04.128& N1& 22.80& 0.74&  GMOS& 0.968& $\circ$& $\times$& $\times$ \\
70399543388149456& 70399543388173932& 16:03:48.794& +43:41:15.429& N1& 24.05& 0.88&  GMOS& 0.886& $\circ$& $\times$& $\times$ \\
70399543388174650& 70399543388174469& 16:03:55.586& +43:41:37.081& N1& 22.03& 0.91&  GMOS& 0.881& $\circ$& $\circ$& $\times$ \\
70399543388174704& 70399543388174546& 16:03:53.369& +43:41:35.007& N1& 22.78& 1.33&  GMOS& 0.977& $\circ$& $\times$& $\times$ \\
70399543388175703& 70399543388175368& 16:03:49.671& +43:42:18.209& N1& 23.04& 0.94& FOCAS& 0.983& $\circ$& $\times$& $\times$ \\
70399543388175737& 70399543388175424& 16:03:52.191& +43:42:20.824& N1& 21.12& 1.22& FOCAS& 0.977& $\circ$& $\circ$& $\times$ \\
70399543388151060& 70399543388175950& 16:03:38.739& +43:42:49.163& N1& 22.60& 0.59&  GMOS& 0.978& $\circ$& $\circ$& $\circ$ \\
70399543388176189& 70399543388175958& 16:03:53.693& +43:42:46.050& N1& 21.95& 1.02& FOCAS& 0.919& $\circ$& $\times$& $\times$ \\
70399543388176413& 70399543388176183& 16:03:42.241& +43:43:07.711& N1& 23.05& 1.10&  GMOS& 0.964& $\circ$& $\times$& $\times$ \\
70399543388176687& 70399543388176496& 16:03:55.607& +43:43:17.332& N1& 22.91& 1.11&  GMOS& 0.977& $\circ$& $\times$& $\times$ \\
70399543388176843& 70399543388176655& 16:03:46.502& +43:43:26.261& N1& 22.81& 0.69& FOCAS& 0.927& $\circ$& $\circ$& $\circ$ \\
70399543388177267& 70399543388177159& 16:03:59.120& +43:43:48.143& N1& 21.40& 1.10& FOCAS& 0.878& $\circ$& $\circ$& $\times$ \\
70399543388177396& 70399543388177276& 16:03:51.165& +43:44:00.382& N1& 23.16& 0.99&  GMOS& 0.899& $\circ$& $\circ$& $\times$ \\
70399543388177591& 70399543388177499& 16:04:02.434& +43:44:04.793& N1& 22.41& 1.07& FOCAS& 0.886& $\circ$& $\circ$& $\times$ \\
70399543388178430& 70399543388178480& 16:03:57.669& +43:44:41.039& N1& 23.46& 1.02& FOCAS& 0.962& $\circ$& $\times$& $\times$ \\
70399543388178966& 70399543388179016& 16:03:42.631& +43:45:05.567& N1& 21.44& 1.02&  GMOS& 0.951& $\circ$& $\times$& $\circ$ \\
70399543388179063& 70399543388179128& 16:03:55.791& +43:45:10.326& N1& 22.88& 0.84&  GMOS& 1.012& $\circ$& ---& --- \\
70399543388179264& 70399543388179385& 16:03:55.148& +43:45:19.152& N1& 22.82& 0.82&  GMOS& 0.881& $\circ$& $\circ$& $\times$ \\
70399680827104655& 70399680827105070& 16:03:36.322& +43:43:11.819& N1& 23.62& 0.99&  GMOS& 0.951& $\circ$& $\times$& $\circ$ \\
70405040946312845& 70405040946316403& 16:04:06.637& +43:44:42.294& N1& 22.04& 0.82& FOCAS& 0.884& $\circ$& $\circ$& $\circ$ \\
70405040946312828& 70405040946316492& 16:04:04.500& +43:44:46.922& N1& 19.96& 1.35& FOCAS& 0.868& $\circ$& $\circ$& $\times$ \\
70405040946317759& 70405040946317910& 16:04:07.673& +43:45:37.973& N1& 22.18& 0.91& FOCAS& 0.890& $\circ$& $\circ$& $\circ$ \\
\end{longtable}
\end{landscape}

\normalsize
\twocolumn

\end{document}